\documentclass[fleqn,usenatbib]{mnras}

\usepackage{newtxtext,newtxmath}

\usepackage[T1]{fontenc}
\usepackage{ae,aecompl}

\usepackage{subcaption}

\usepackage{xcolor}
\usepackage{graphicx}	
\usepackage{amsmath}	

\usepackage{epsfig}
\usepackage{gensymb}
\usepackage{rotating}
\usepackage{color}
\usepackage{tabularx}
\usepackage{multirow}
\usepackage{graphics}
\usepackage{epsfig}
\usepackage{epstopdf}
\usepackage[para,online,flushleft]{threeparttable}
\usepackage[ampersand]{easylist}
\usepackage{hyperref}
\hypersetup{
    colorlinks=true,
    linkcolor=blue,
    filecolor=blue,      
    urlcolor=blue,
    citecolor=blue,
    filecolor=blue,
}
\title[Optical and HI observations of G25.1$-$2.3]{Optical observations and atomic environment of supernova remnant G25.1$-$2.3}

\author[E.~Aktekin et al.]{{E.~Aktekin$^{1}$\thanks{{\color {blue}corresponding author: ebrucaliskan@sdu.edu.tr} (EA)},
H.~Bak{\i}\c{s}$^{2}$,
V.~Bak{\i}\c{s}$^{2}$,
Y.~Asano$^{3}$, 
H.~Sano$^{3}$, 
Y.~Fukui$^{3,4}$ and
A.~Sezer$^{5}$}\\
$^{1}$Department of Physics, S\"{u}leyman Demirel University, Isparta 32000, T\"{u}rkiye \\
$^{2}$Department of Space Sciences and Technologies, Akdeniz University, Antalya 07058, T\"{u}rkiye\\
$^{3}$Faculty of Engineering, Gifu University, 1-1 Yanagido, Gifu 501-1193, Japan\\
$^{4}$Department of Physics, Nagoya University, Furo-cho, Chikusa-ku, Nagoya 464-8601, Japan\\
$^{5}$Department of Computer Engineering, Avrasya University, Trabzon 61250, T\"{u}rkiye\\
}

\date{Accepted XXX. Received YYY; in original form ZZZ}

\pubyear{2026}

\begin{document}
\label{firstpage}
\pagerange{\pageref{firstpage}--\pageref{lastpage}}
\maketitle

\begin{abstract}
The supernova remnant (SNR) G25.1$-$2.3 was identified in the radio band during the Sino-German $\lambda$6~cm survey of the Galactic plane. We present a detailed investigation of the optical, H\,{\sc i}, and CO emission towards the G25.1$-$2.3 to better understand its characteristics and environment.  
In this study, optical spectroscopic data of the remnant and its environment have been analysed for the first time, providing new insights into their emission properties. The Large Sky Area Multi-Object Fiber Spectroscopic Telescope (LAMOST) and 1.5-m Russian-Turkish Telescope (RTT150) data show variations across the observed regions, with [S\,{\sc ii}]/H$\alpha$ ranging from $0.16$ to $0.83$.  We identified shock-heated gas in the northern and southern regions and several photoionized regions around the SNR based on their [S\,{\sc ii}]/H$\alpha$ ratios derived from spectra. The [S\,{\sc ii}]$\lambda$6716/$\lambda$6731 ratio observed in the northern region suggests electron densities ($n_{\rm e}$) ranging from 120 to 1030 cm$^{-3}$, whereas the southern regions show higher values, between 490 and 4500 cm$^{-3}$. The variations in the observed H$\alpha$/H$\beta$ ratios indicate significant differences in extinction across the regions. H$\alpha$ images obtained using the 1-m Turkish Telescope (T100) reveal optical emission in the northern and southern, characterized by filamentary and diffuse structures. We newly found a hole-like distribution of H\,{\sc i}, whose spatial extent is roughly consistent with the diameter of the SNR. Based on radio data, we examine the evolutionary stage of G25.1$-$2.3 using the surface brightness–diameter ($\Sigma-D$) relation and the equipartition method.
\end{abstract}




\section{Introduction}   
Over the past two decades, numerous new Galactic supernova remnants (SNRs) and candidates have been discovered through radio continuum surveys (e.g. \citealt{Gr14, Ko17, An17, Ball2023, Ball2025, Filipovic2025, Anderson2025}). Follow-up optical observations, combined with X-ray, gamma-ray, H\,{\sc i}, and CO data, provide valuable information about the physical characteristics of the SNRs and their surrounding medium (e.g.  G118.4+37.0: \citealt{Arias2022, Araya2023, Greco2025}, G107.7$-$5.1: \citealt{Fesen2024, Araya2024}, Perun: \citealt{Smeaton2024}, Ancora: \citealt{Filipovic2023, Burger-Scheidlin2024}, Diprotodon: \citealt{Stupar2011, Filipovic2024}, G206.7+5.9: \citealt{Gao22, Bakis2025a}).

G25.1$-$2.3, with an angular size of approximately 80 $\times$ 30 arcmin$^{2}$ (RA=18$^{\rmn{h}}$45$^{\rmn{m}}$09$^{\rmn{s}}$; Dec.= $-07$$\degr$59$\arcmin$42$\arcsec$), was identified as an SNR by \citet{Ga11}, based on a strong southern shell observed in the radio map from the Sino-German $\lambda$6~cm polarization survey of the Galactic plane. Further observations with the Effelsberg $\lambda$11 and $\lambda$21~cm radio continuum, along with polarization measurements, confirmed its shell-type morphology. The radio spectra show non-thermal emission characteristics, with a spectral index for the shell of $\alpha = -0.49 \pm 0.13$ ($S$ $\propto$ $\nu ^{\alpha}$). They compared the radio map at $\lambda$11~cm of G25.1$-$2.3 with the Southern H-Alpha Sky Survey Atlas (SHASSA; \citealt{Ga01}) H$\alpha$ image. A prominent H$\alpha$ emission feature was observed north of the SNR shell, while the emission within the shell itself is comparatively faint. Consequently, they suggested that the H$\alpha$ emission in this direction is probably not physically associated with G25.1$-$2.3. They also investigated an associated H\,{\sc i} cavity and identified one in the velocity range of 37.1$-$41.2~km~s$^{-1}$ within the area of the SNR. They found that the cavity's shape closely resembles that of the SNR, suggesting a possible connection between them. 

Based on the association of the SNR with an H\,{\sc i} cavity, \citet{Ga11} derived a kinematic distance of approximately 2.9~kpc. 
Further kinematic distance estimates were proposed by \citet{Wa20} as $d \sim 3.5$~kpc, and by \citet{Ra22} as $d \sim 2.7$~kpc. A distance range of approximately 2.6$-$3.9~kpc has previously been suggested based on the surface brightness-diameter ($\Sigma-D$) relation \citep{Pavlovic2014}. More recently, \citet{Urosevic2025} estimated a distance of about 2.2$-$2.5~kpc using the updated $\Sigma-D$ relation.

The SNR G25.1$-$2.3 has so far been detected only at radio frequencies, and its physical properties, as well as the characteristics of its surrounding environment, remain unexplored in other wavelengths. In this work, we present an optical study of the SNR G25.1$-$2.3, aimed at determining its physical characteristics. We analysed optical spectra towards G25.1$-$2.3 from the Large Sky Area Multi-Object Fiber Spectroscopic Telescope (LAMOST)\footnote{\url{https://www.lamost.org/}}  Data Release 9 (DR9) and 1.5-m Russian-Turkish Telescope (RTT150)\footnote{\url{https://trgozlemevleri.gov.tr/teleskoplar/antalya/rtt150}}. Additionally, we conducted optical imaging observations of the northern and southern regions using the 1-m Turkish Telescope (T100)\footnote{\url{https://trgozlemevleri.gov.tr/teleskoplar/antalya/tug100}} and 1.5-m RTT150 telescopes at T\"{u}rkiye National Observatories\footnote{\url{https://trgozlemevleri.gov.tr/}}. To further explore the interaction between the SNR and its environment, we also examined H\,{\sc i} and CO data. Section~\ref{sec:obs} describes the observations and data reduction process, followed by Section~\ref{sec:analysis}, which presents the analysis and corresponding results. Section~\ref{sec:discuss} discusses the physical properties of G25.1$-$2.3 and provides an interpretation of the results, and Section~\ref{sec:conc} concludes the study.


\section{Observations and data reduction}
\label{sec:obs}
\subsection{Optical imaging data}      
We conducted H$\alpha$ ($\lambda_{\rm c}$=656.0~nm and FWHM=10.8~nm) imaging of G25.1$-$2.3 using the 1.0~m fully automated Ritchey-Chrétien T100 telescope. 
The CCD camera features a resolution of $4096 \times 4096$ pixels, offering a field of view (FoV) of $21.5 \times 21.5$ arcmin$^2$. We observed five regions (namely S1, S2, S3, NW, and N) in the southern and northern parts of G25.1$-$2.3. In addition to the T100 telescope, the 1.5-m RTT150 telescope was also used for  H$\alpha$ ($\lambda_{\rm c}$=656.3~nm and FWHM=5~nm) imaging of S3 filament. The CCD camera is composed of 2048 $\times$ 2048 pixels providing a FoV of 11.1 $\times$ 11.1 arcmin$^2$. A summary of the imaging observation details is provided in Table~\ref{tab:Table1}.

We processed the imaging data using {\sc iraf} (Image Reduction and Analysis Facility)\footnote{\url{https://iraf-community.github.io/}}, starting with bias subtraction and flat-field corrections. Next, the images were processed to remove defective pixels and cosmic-ray artifacts from the CCD frames. After these steps, the continuum emission was subtracted from each image taken with the H$\alpha$ filter using a continuum filter ($\lambda_{\rm c}$=644.6~nm and FWHM=13~nm for both T100 and RTT150). Since the continuum image was obtained with a shorter exposure time, a scale factor ($k$) was determined prior to the subtraction. For this purpose, aperture photometry was performed on at least ten non-saturated field stars within the H$\alpha$ and H$\alpha_{\mathrm{cont}}$ images. The flux for each star was measured, and the scale factor was obtained as the average of the flux ratios between the H$\alpha$ and H$\alpha_{\mathrm{cont}}$ images, defined as

\begin{equation}
k = \left\langle \frac{I_{\mathrm{H\alpha}}}{I_{\mathrm{H\alpha,cont}}} \right\rangle
\end{equation}
The continuum-subtracted image was then produced using
\begin{equation}
I_{\mathrm{cont-sub}} = I_{\mathrm{H\alpha}} - k \times I_{\mathrm{H\alpha,cont}}
\end{equation}

This effectively removed the stellar continuum, allowing for a clearer detection of the full spatial extent of the faint diffuse emission.

\begin{table*}
 \caption{The main characteristics of the H$\alpha$ imaging observations of G25.1$-$2.3 with the T100 and RTT150.}
 \label{tab:Table1}
 \begin{threeparttable}
 \begin{tabular}{@{}p{1.8cm}p{1.5cm}p{2.1cm}p{2.1cm}p{2.0cm}p{2.3cm}p{1.5cm}@{}}
 \hline
Region ID & RA (J2000) & Dec. (J2000) &  Average seeing &   Exposure & Observation & Telescope  \\
          &  (h m s)   & ($\degr$ $\arcmin$ $\arcsec$) &      (arcsec) &   (s)   & date & \\
\hline
S1   &  18 46 27.0  & -07 34 12.3    &    1.3  &  3 $\times$ 1200    &  2025 Jul 23  & T100  \\
S2   &  18 46 04.9  & -07 46 35.1    &    1.3  &  3 $\times$ 1200    &  2025 Jul 23  &  T100 \\
S3   &  18 45 14.9  & -07 58 52.3    &    1.3  &  3 $\times$ 1200    &  2025 Jul 23  & T100  \\
NW   &  18 42 45.6  & -07 55 50.0    &    1.4  &  3 $\times$ 1200    &  2025 Jul 22  & T100  \\
N    &  18 44 07.1  & -07 19 19.6    &    1.4  &  3 $\times$ 1200    & 2025 Jul 22   & T100  \\
S3   & 18 45 25.9   & -07 49 31.3    &    1.4  &  3 $\times$ 1200    &  2025 Jul 31  & RTT150  \\

 \hline
\end{tabular}
\begin{tablenotes}
\item {{\it Note.} Following each observation with the  H$\alpha$ filter, the same region was subsequently observed using a continuum filter to subtract the background. The exposure time was 3 $\times$ 400 s for each continuum image.}\\
\end{tablenotes}
\end{threeparttable}
\end{table*}

\subsection{LAMOST spectral data}
LAMOST is a 4-m quasi-meridian reflecting Schmidt telescope with a FoV spanning 5$\degree$ in diameter \citep{Cu12}. It operates in two modes: low- and medium-resolution. The medium-resolution mode covers 3700--5900 {\AA} (blue) and 5700--9000 {\AA} (red) channels. We used medium-resolution spectra with a resolution of $R$ $\sim$ 7500. The slit (2/3) has a diameter of 2.2  arcsec. The observations were made on May 22, 2019, with an exposure time of 1200 s. The \texttt{pylamost.py}\footnote{\url{https://github.com/fandongwei/pylamost}} code was used to access the LAMOST data.
      
\subsection{RTT150 spectral data}
Long-slit spectra of four positions (S2, S3, NW, and N) of the remnant were obtained with the TUG Faint Object
Spectrograph and Camera (TFOSC) installed at the 1.5-m RTT150 telescope. The setup employed a grism 15 and a slit measuring 2.38 arcsec $\times$ 234 arcsec, covering 3230$-$9120 {\AA} with a resolution of $R$ $\sim$ 749. Log of spectroscopic observations with RTT150 is provided in Table~\ref{tab:Table2}.

Data reduction was carried out with  {\sc iraf}, using an iron–argon lamp for wavelength calibration and the BD+28 4211 standard star \citep{Oke1990} for flux calibration. 

\begin{table*}
\centering
 \caption{Spectroscopic observations of G25.1$-$2.3 with the RTT150.}
 \label{tab:Table2}
 \begin{tabular}{@{}p{1.8cm}p{1.8cm}p{1.8cm}p{1.8cm}p{1.8cm}@{}}
 \hline
Slit ID &    RA (J2000) & Dec. (J2000)                   & Exposure & Observation  \\
   &  (h m s)      & ($\degr$ $\arcmin$ $\arcsec$)  &  (s)     & date \\
\hline
S2 &  18 46 12.8 &  -07 47 55.4     & $3 \times 1200$    & 2025 Sep 21    \\
S3 &  18 45 20.9 & -07 48 55.9     & $3 \times 1200$    & 2025 Aug 24    \\
NW &  18 42 39.3 & -07 55 32.1     & $3 \times 1200$    & 2025 Aug 27    \\
N &   18 44 37.2 & -07 17 12.4     & $3 \times 1200$    & 2025 Aug 29    \\
\hline
\end{tabular}
\end{table*}

\section{Analysis and Results}
\label{sec:analysis}
\subsection{Optical images}
\begin{figure*}
\includegraphics[angle=0, width=12cm]{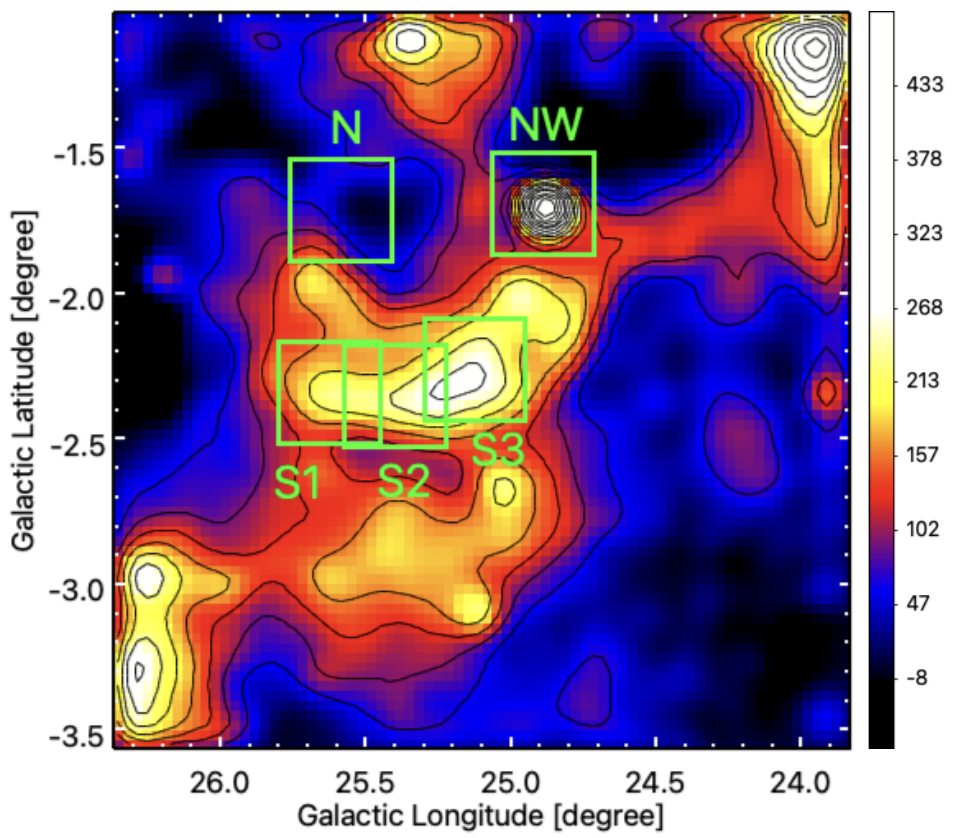}
\caption{The $\lambda$11~cm radio continuum image \citep{Re90} of the area around G25.1$-$2.3. The radio continuum contours scale linearly from 60 to 488~mJy~beam$^{-1}$. We show five regions (S1, S2, S3, NW, and N) observed with the 1-m T100 telescope, marked with boxes.}  
\label{fig:figure1}
\end{figure*}

In Fig.~\ref{fig:figure1}, we display the Effelsberg radio continuum image at $\lambda$11~cm with radio contours \citep{Re90}. The five regions (S1, S2, S3, NW, and N) observed with the 1-m T100 telescope are marked with boxes.

\begin{figure*}
\includegraphics[angle=0, width=16cm]{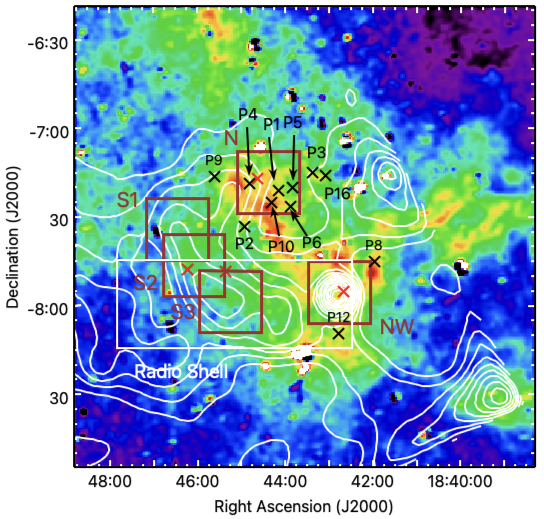}
\caption{The continuum-corrected SHASSA \citep{Ga01} H$\alpha$ image ($2.6 \times 2.6$ deg$^{2}$) of the SNR's vicinity, highlighting the radio size of the SNR (white box; $80 \times 30$~arcmin$^{2}$) from \citet{Ga11}, the regions observed by the 1-m T100 telescope (brown boxes; $21 \times 21$~arcmin$^{2}$), and the locations where the LAMOST and RTT150 spectra were extracted, marked with black and red crosses, respectively. The overlaid  $\lambda$11~cm radio continuum contours \citep{Re90} scale linearly from 60 to 488~mJy~beam$^{-1}$.} 
\label{fig:figure2a}
\end{figure*}

In Fig.~\ref{fig:figure2a}, we provide a $2.6 \times 2.6$ deg$^{2}$ view  of G25.1$-$2.3 and its environment using the continuum-corrected SHASSA \citep{Ga01} H$\alpha$ image. The regions observed by the 1-m T100 telescope are highlighted with brown boxes, while the radio extent of the SNR is marked by a white box, which spans $80 \times 30$ arcmin$^{2}$ \citep{Ga11}. Additionally, the positions of the extracted LAMOST and RTT150 spectra are indicated by black and red crosses, respectively. The SHASSA H$\alpha$ image reveals prominent filamentary structures located to the north of the SNR shell. 

\begin{figure*}
\includegraphics[angle=0, width=6.6cm]{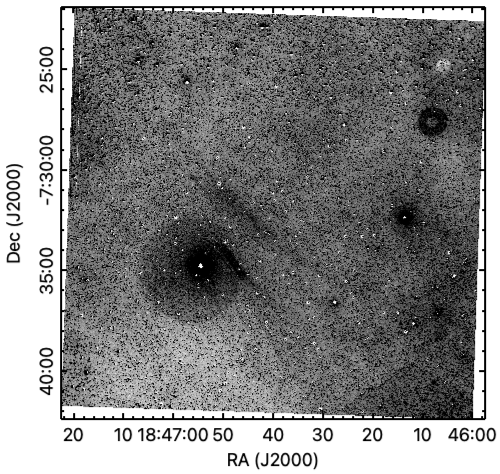}
\includegraphics[angle=0, width=6.5cm]{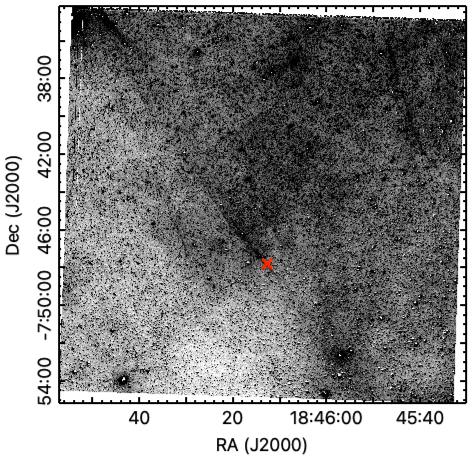}
\includegraphics[angle=0, width=6.5cm]{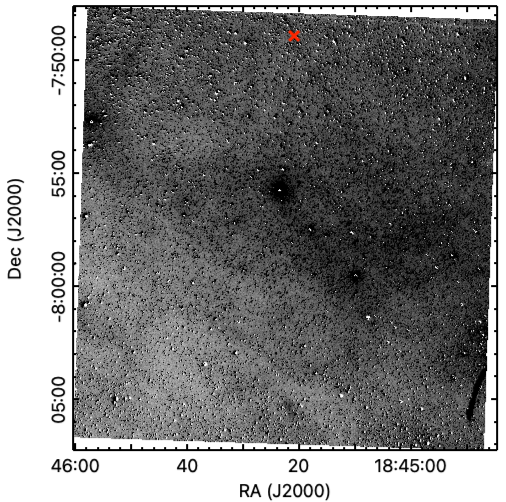}
\includegraphics[angle=0, width=6.7cm]{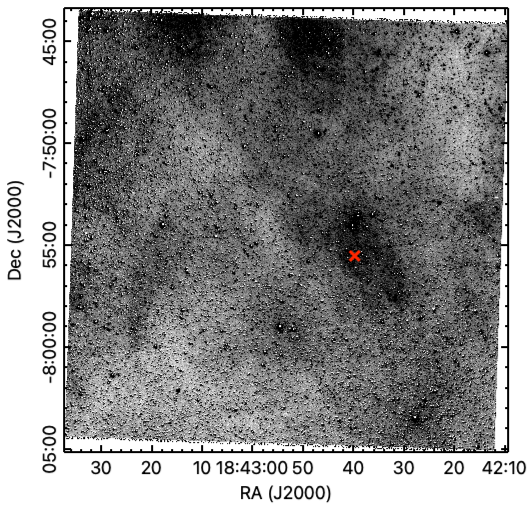}
\includegraphics[angle=0, width=6.5cm]{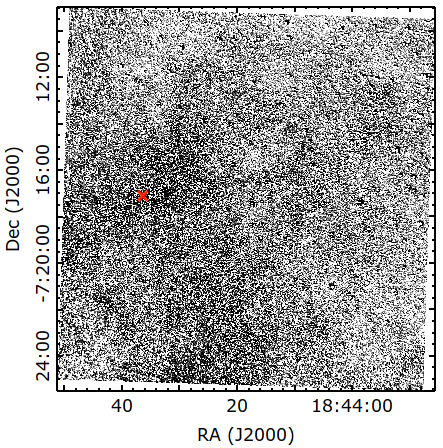}
\includegraphics[angle=0, width=6.7cm]{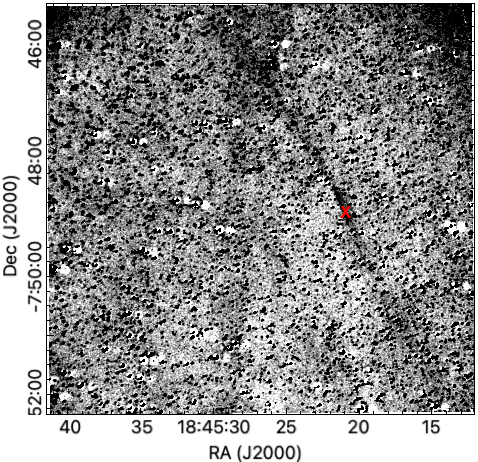}
\caption{The continuum-subtracted H$\alpha$ images for S1, S2, S3, NW, and N regions with the 1-m T100 telescope, starting from the top left, respectively. The bottom-right panel shows a continuum-subtracted H$\alpha$ image focusing on the filament located in the S3 region, obtained with the 1.5-m RTT150 telescope.  The slit positions from the RTT150 spectroscopic observations are shown as crosses, and their central coordinates are listed in Table~\ref{tab:Table2}.}
\label{fig:figure2b}
\end{figure*}

In Fig.~\ref{fig:figure2b}, we show the continuum-subtracted H$\alpha$ images of the S1, S2, S3, NW, and N regions of the SNR, which were obtained using the 1-m T100 telescope. The filament detected in the S3 region with T100 was also observed with the RTT150 telescope to see its structure in greater detail. The continuum-subtracted H$\alpha$ image of the filament is presented in the bottom-right panel of Fig.~\ref{fig:figure2b}.

\subsection{Optical spectra} 
 \label{sec:discuss-spectra}
To examine the spectral characteristics, we obtained LAMOST and RTT150 spectra from various regions within the SNR and its surroundings (see  Figs~\ref{fig:figure2a} and \ref{fig:figure2b}). For the spectral modeling, fitting, and uncertainties, we followed the methods detailed in \citet{Bakis2025a}. 

\begin{figure*}
\includegraphics[angle=0, width=17cm]{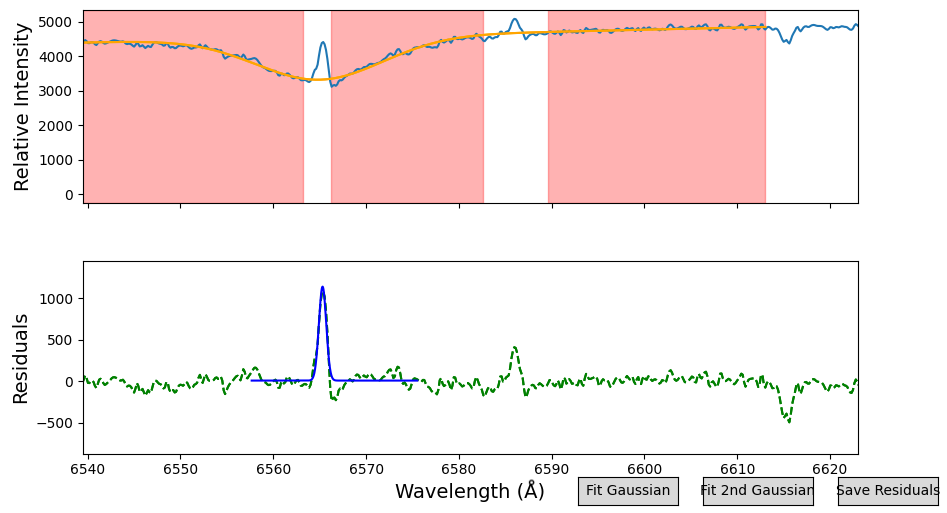}
\caption{Top panel: Example spectrum including the H$\alpha$ stellar absorption feature. Bottom panel: Residual spectrum after Gaussian fit. The yellow line shows the Gaussian fit to the stellar absorption. Pink areas refer to fitted sampling regions.}
\label{fig:figure3a}
\end{figure*}

The LAMOST spectra towards G25.1$-$2.3 exhibit distinct spectral features, including (I) Shock-heated region, (II) Shock-heated region + stellar, (III)  Photoionized region, and (IV) Photoionized region + stellar. The H$\alpha$ stellar absorption features in the spectra are corrected using a Gaussian fit. An example spectrum is shown in Fig.~\ref{fig:figure3a}. The spectra in the ranges $6540-6600$ and $6710-6750$ {\AA} are shown in Figs~\ref{fig:figure3} and \ref{fig:figure4}. In Figs~\ref{FigA1}-\ref{FigA3}, we provide additional spectra for these wavelength ranges, as well as for $6290-6370$ {\AA}, including  [O\,{\sc i}] $\lambda$$\lambda$6300, 6363 emission lines. The Balmer line H$\alpha$ $\lambda$6563, and forbidden lines [O\,{\sc i}] $\lambda$$\lambda$6300, 6363, [N\,{\sc ii}] $\lambda$6584 and [S\,{\sc ii}] $\lambda$$\lambda$6716, 6731 are seen in the LAMOST spectra. We list emission line fluxes relative to H$\alpha$ in Table~\ref{tab:Table3}. We also calculate the line ratios of [S\,{\sc ii}]/H$\alpha$ and [N\,{\sc ii}]/H$\alpha$ with errors (see Table~\ref{tab:Table3}). We provide the general characteristics and spectral classification of the selected regions extracted from the LAMOST data in Table~\ref{tab:Table4}. 
 
Additionally, spectra were obtained from regions S2, S3, NW, and N using the RTT150. Fig.~\ref{fig:figure4b} presents the observed spectra, while Table~\ref{tab:Table4b} lists the emission-line fluxes relative to H$\alpha$ along with the line ratios.

\begin{figure*}
\includegraphics[angle=0, width=8.0cm]{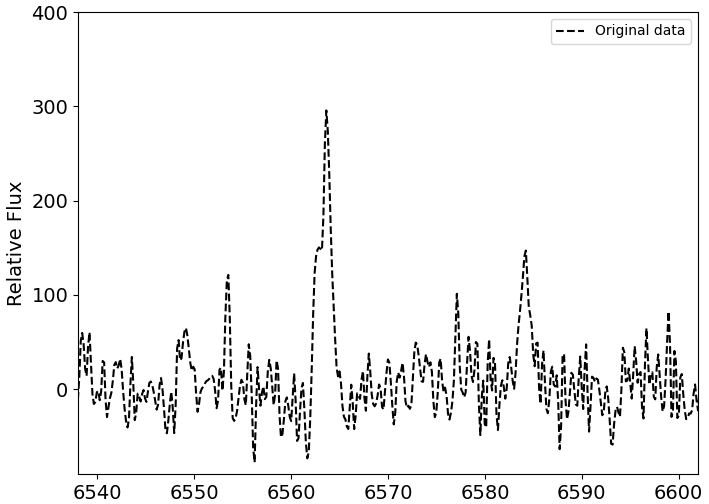}
\includegraphics[angle=0, width=8.0cm]{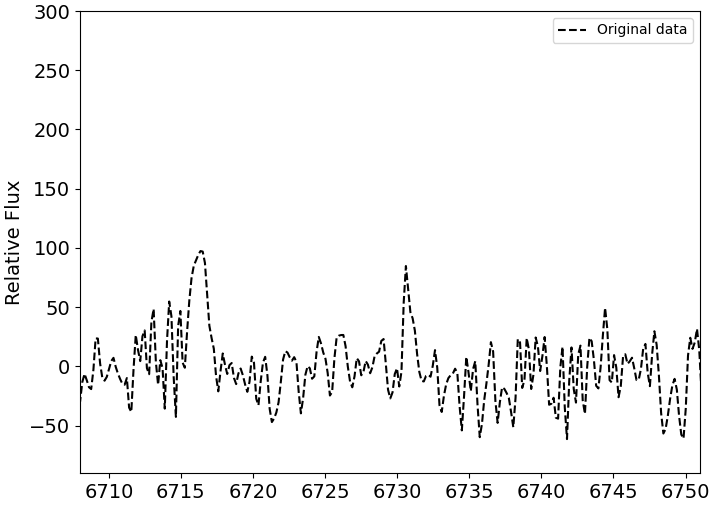}
\includegraphics[angle=0, width=8.0cm]{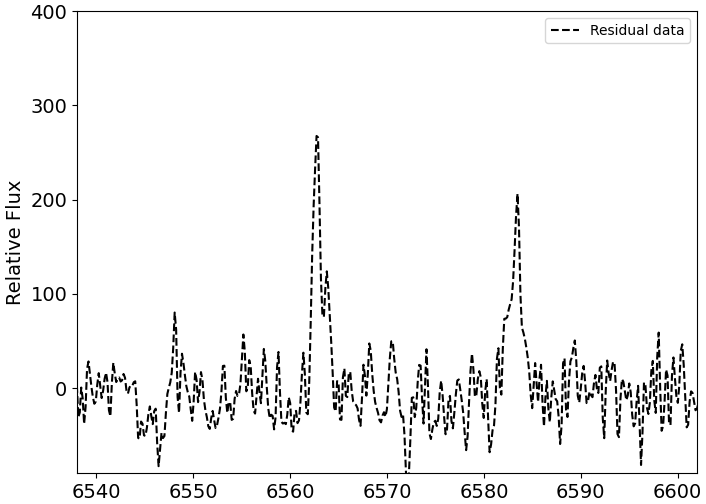}
\includegraphics[angle=0, width=8.0cm]{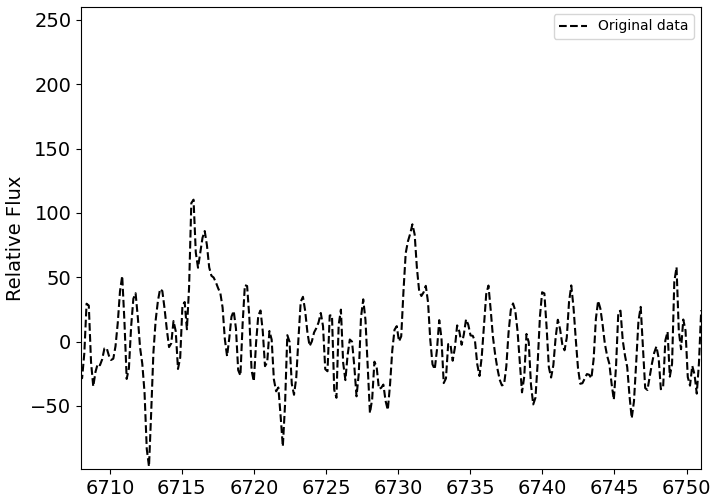}
\includegraphics[angle=0, width=8.0cm]{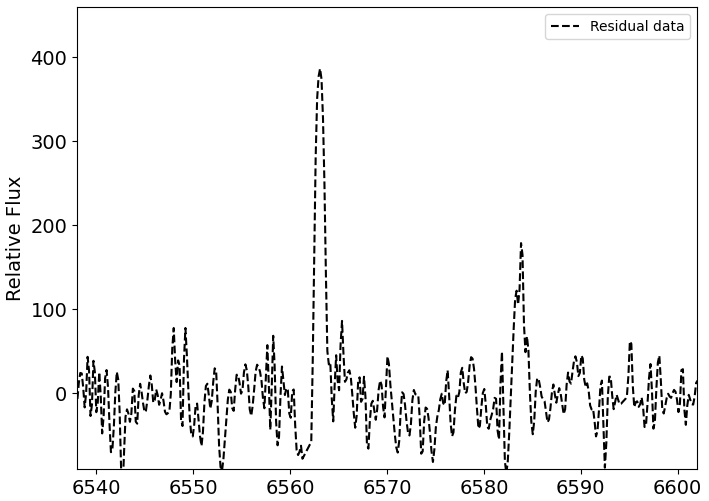}
\includegraphics[angle=0, width=8.0cm]{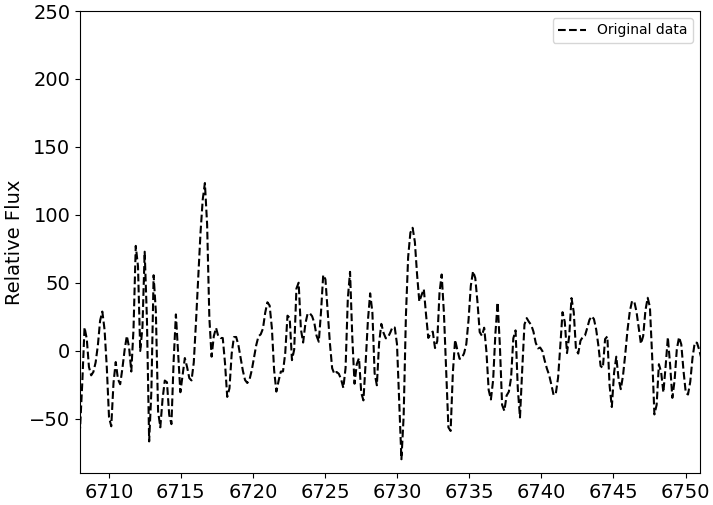}
\includegraphics[angle=0, width=8.0cm]{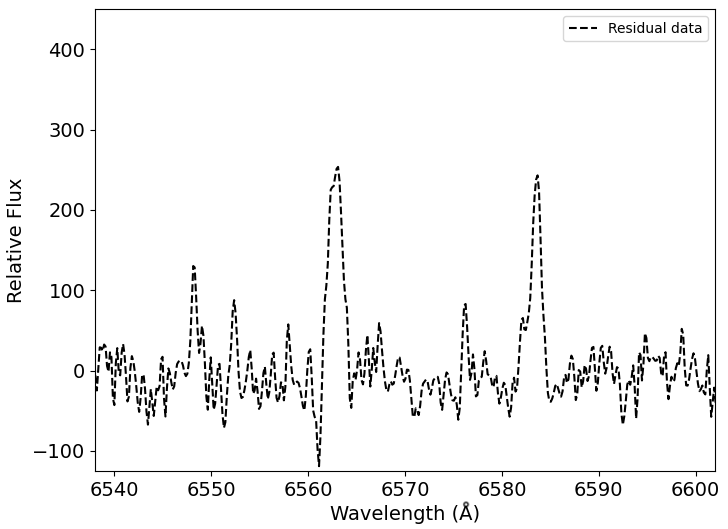}
\includegraphics[angle=0, width=8.0cm]{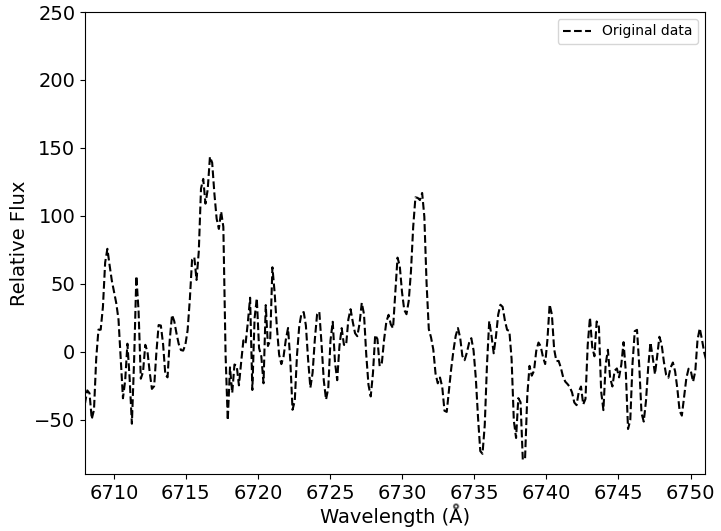}
\caption{The LAMOST spectra ($6540-6600$ and $6710-6750$ {\AA}) for the P1, P2, P3, and P4 positions. The residual data in the upper right corner refer to the spectrum after the stellar line is removed (see Fig.~\ref{fig:figure3a}). The original data refer to the spectrum with detrending only. This is used in the same sense in all figures. All the spectra in these images and the ones that follow are presented in a clockwise order, starting from the top-left.} 
\label{fig:figure3}
\end{figure*}

\begin{figure*}
\includegraphics[angle=0, width=8.0cm]{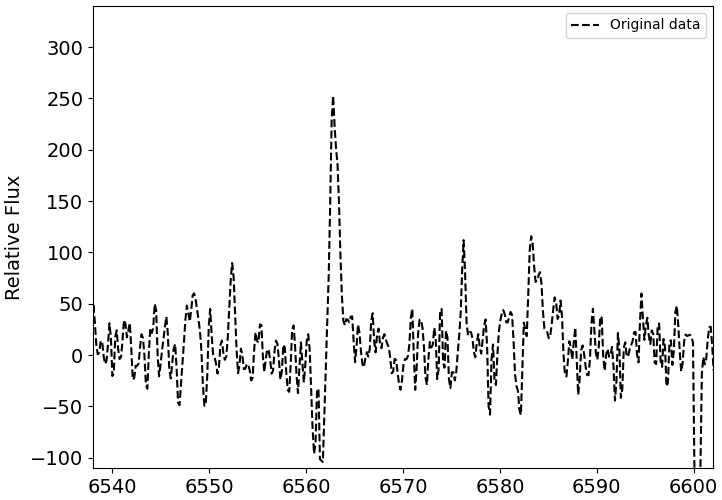}
\includegraphics[angle=0, width=8.0cm]{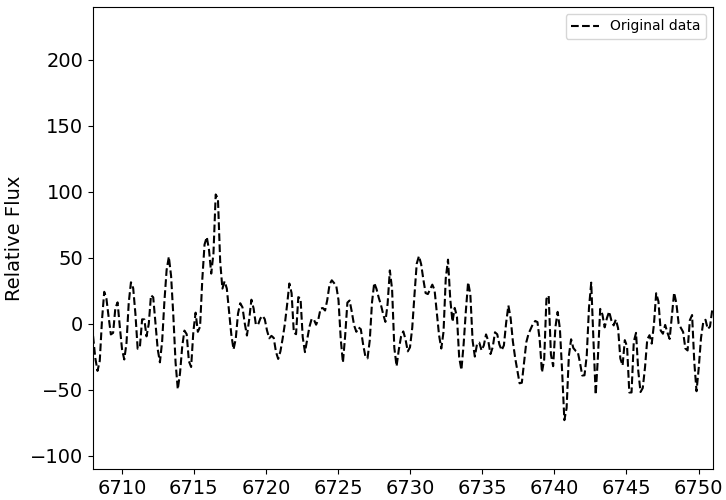}
\includegraphics[angle=0, width=8.0cm]{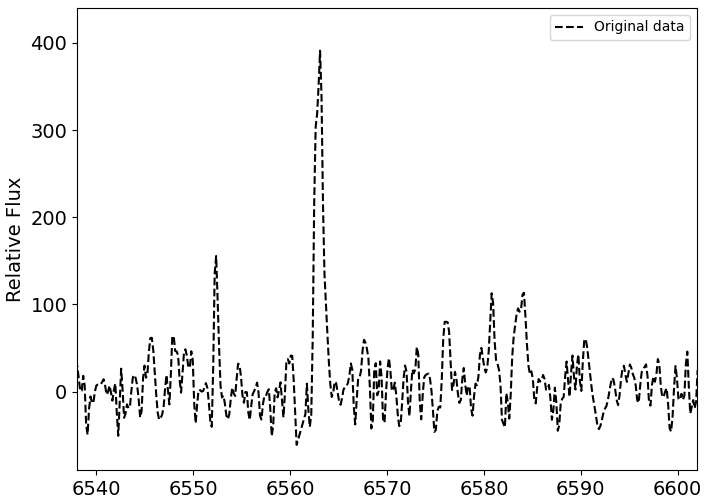}
\includegraphics[angle=0, width=8.0cm]{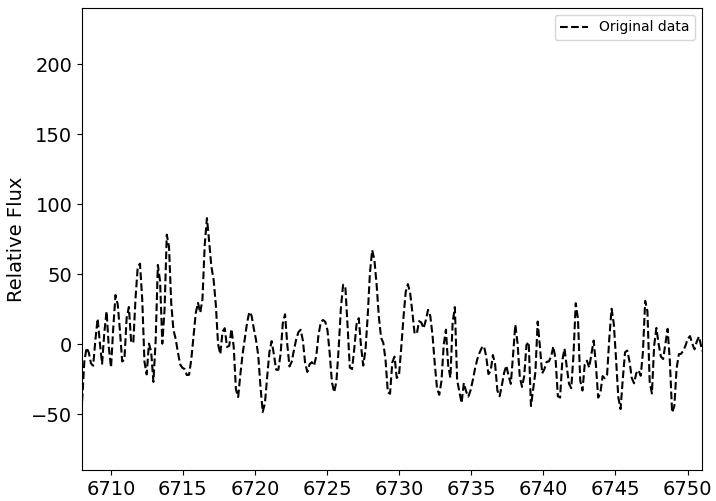}
\includegraphics[angle=0, width=8.0cm]{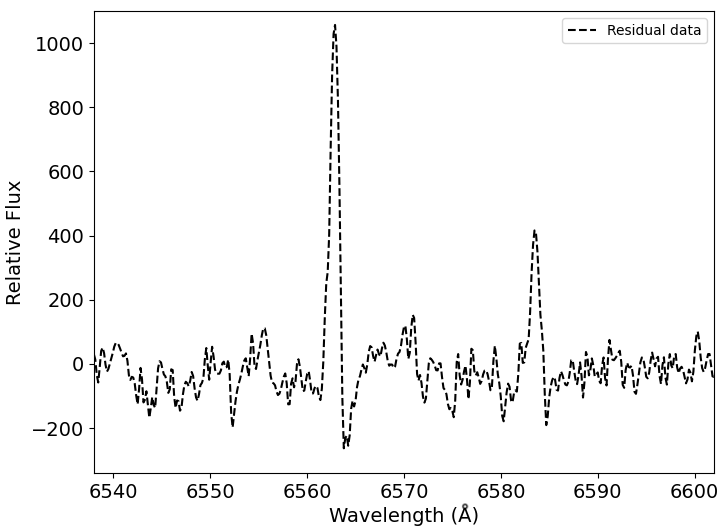}
\includegraphics[angle=0, width=8.0cm]{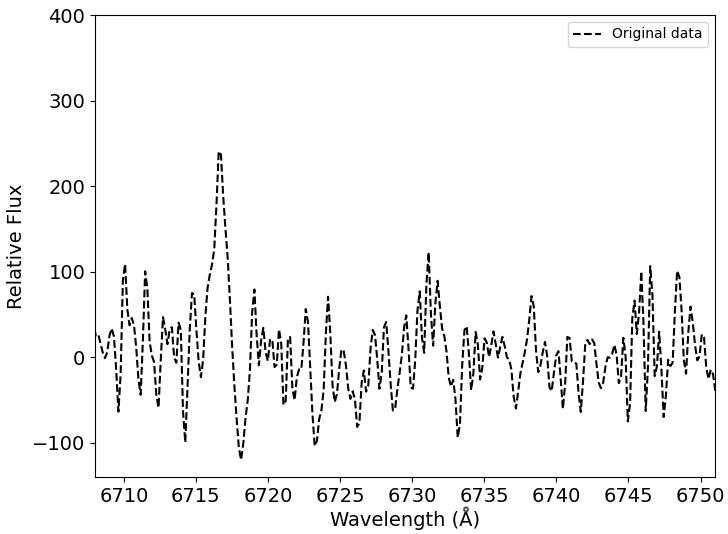}
\caption{Continued from Fig.~\ref{fig:figure3}. The LAMOST spectra ($6540-6600$ and $6710-6750$ {\AA}) for the P5, P6, and P8 positions.}
\label{fig:figure4}
\end{figure*}

\begin{figure*}
\includegraphics[angle=0, width=8.0cm]{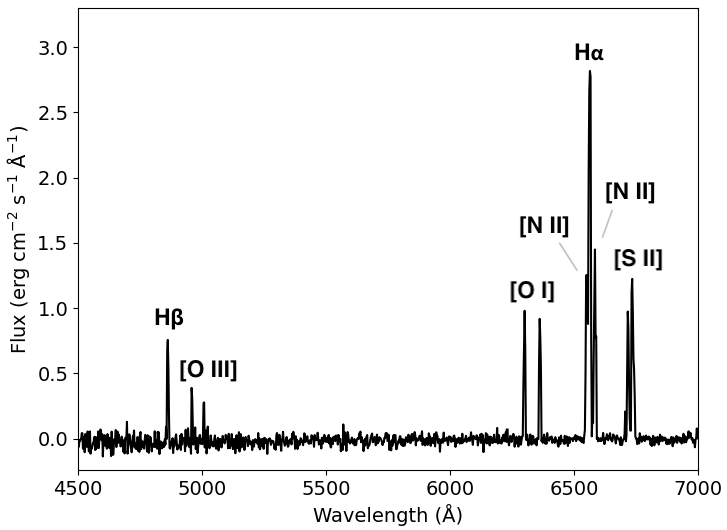}
\includegraphics[angle=0, width=8.0cm]{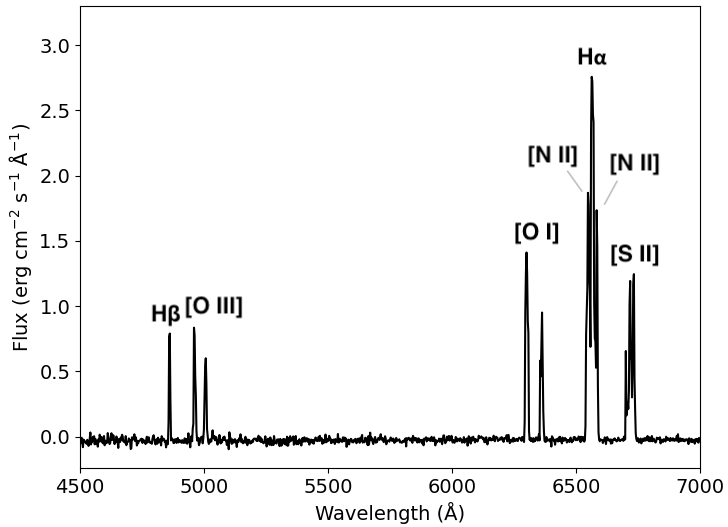}
\includegraphics[angle=0, width=8.0cm]{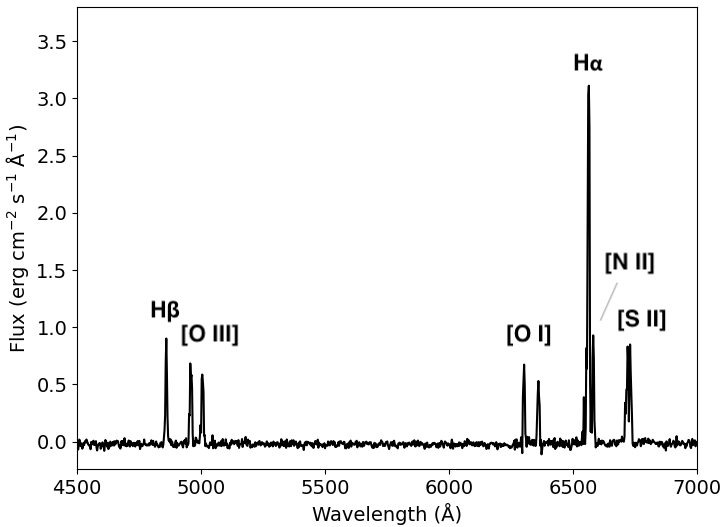}
\includegraphics[angle=0, width=8.0cm]{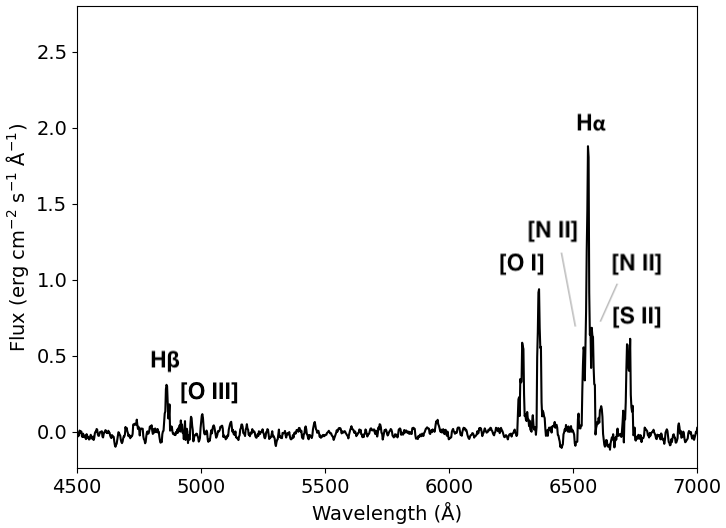}
\caption{The long-slit spectra obtained with the  RTT150 telescope for the S2, S3, NW, and N positions, starting from the top left, respectively. The spectra cover $4500-7000$ {\AA}, with fluxes expressed in units of $10^{-16}$ erg cm$^{-2}$ s$^{-1}$ {\AA}$^{-1}$.}
\label{fig:figure4b}
\end{figure*}

\begin{table*}
\centering
\caption{Relative line fluxes from the LAMOST spectra taken from different locations with the 1-$\sigma$ errors. The measured [S\,{\sc ii}]/H$\alpha$ and  [N\,{\sc ii}]/H$\alpha$ emission line ratios are also listed.}
\label{tab:Table3}
 \begin{tabular}{@{}p{3.5cm}p{1.8cm}p{1.8cm}p{1.8cm}p{1.8cm}@{}}
 \hline
 	
 	& P1 &	P2 	&	P3  &	P4 		 \\[0.5 ex]
\hline		
										
$[$O$\,${\sc i}$]$ $\lambda$6300 	    	& 50$\pm$6 &		$...$         & 23$\pm$5    &	54$\pm$6  			 \\
$[$O$\,${\sc i}$]$ $\lambda$6363 	    	&   31$\pm$7   &		$...$         &   $...$  &		39$\pm$4 		 \\

$[$N$\,${\sc ii}$]$ $\lambda$6548 		& $...$ &  36$\pm$10 	&  $...$  	 &  $...$  \\
																		
H$\alpha$ $\lambda$6563 	             	& 100$\pm$5 &	100$\pm$6	& 100$\pm$4	     &	 100$\pm$5   	  \\
												
$[$N$\,${\sc ii}$]$ $\lambda$6584 		&  52$\pm$4 &  68$\pm$5 	&  38$\pm$4  	 &  135 $\pm$4  \\
												
$[$S$\,${\sc ii}$]$ $\lambda$6716 		& 39$\pm$4  &	37$\pm$5    &  30$\pm$4   &	  46$\pm$6   \\
												
$[$S$\,${\sc ii}$]$ $\lambda$6731 		& 29$\pm$5&	36$\pm$5	&  22$\pm$4    &	37$\pm$7 	\\[0.5 ex]
												
[S\,{\sc ii}]/H$\alpha$ 	                &  0.68$\pm$0.04 &0.73$\pm$0.02  &	0.51$\pm$0.03 	 &	 0.83 $\pm$0.04	  \\[0.5 ex]

[N\,{\sc ii}]/H$\alpha$ 	               &0.52$\pm$0.02 & 1.04$\pm$0.08  &	0.38$\pm$0.02 	 &   1.35 $\pm$0.05 	 \\[0.5 ex]	
\hline
&			P5 &	P6 &	P8	&  P9	 \\[0.5 ex]
\hline												
$[$O$\,${\sc i}$]$ $\lambda$6300 	    	&	56$\pm$5         &	26$\pm$2   &		$...$ &	 $...$	 \\
$[$O$\,${\sc i}$]$ $\lambda$6363 	    	&   28$\pm$6         &	29$\pm$3   &    	$...$	&	$...$	 \\
																		
H$\alpha$ $\lambda$6563 	             	&	100$\pm$5 &	100$\pm$4	&	  	 100$\pm$4 &	100$\pm$3  \\
												
$[$N$\,${\sc ii}$]$ $\lambda$6584 		& 		40$\pm$5 &	29$\pm$2 & 	   40$\pm$3 & 44$\pm$3\\
												
$[$S$\,${\sc ii}$]$ $\lambda$6716 		&	    29$\pm$4   &	19$\pm$3    &		 20$\pm$2   &  24$\pm$6  \\
																								
[S\,{\sc ii}]/H$\alpha$ 	 &			   0.29$\pm$0.06 &	0.19$\pm$0.05 &	   0.20$\pm$0.01	&  0.24 $\pm$0.05 \\[0.5 ex]

[N\,{\sc ii}]/H$\alpha$ 	&	 		  0.40$\pm$0.03 &	0.29$\pm$0.01 &    0.40$\pm$0.01	& 0.44 $\pm$0.02 \\[0.5 ex]	
 \hline
 		&		  	P10  &    P12 & P16 &	 \\[0.5 ex]
\hline												
$[$O$\,${\sc i}$]$ $\lambda$6300 	    	  &	28$\pm$4 	  &	$...$	& 	$...$ &	 \\																		
H$\alpha$ $\lambda$6563 	             	  & 100$\pm$5	&	 100$\pm$5 &  100$\pm$5 & \\
												
$[$N$\,${\sc ii}$]$ $\lambda$6584 		  & 23$\pm$2 &     43$\pm$4 & 45$\pm$5  &  \\
												
$[$S$\,${\sc ii}$]$ $\lambda$6716 		  &  16$\pm$2	&  	$...$ & $...$ &	 \\
																								
[S\,{\sc ii}]/H$\alpha$ 	         	      & 0.16 $\pm$0.02	&	    	$...$ & $...$	& \\[0.5 ex]

[N\,{\sc ii}]/H$\alpha$ 		             & 0.23 $\pm$0.01      & 0.43 $\pm$0.02 & 0.45$\pm$0.03 & \\[0.5 ex]	
\hline
\end{tabular}
\end{table*}

\begin{table*}
\begin{threeparttable}
\centering
 \caption{General characteristics and spectral classification of the selected regions extracted from the LAMOST data.}
 \label{tab:Table4}
 \begin{tabular}{@{}p{1.2cm}p{1.8cm}p{3.0cm}p{3.8cm}@{}}
 \hline
Slit ID & Obs.ID & ~~~~~~~~Position  & Optical Spectral  \\
 &  & ~~~~~~~(RA  ; Dec.) &  Description$^{\ast}$   \\
\hline
P1 &  745201147 & 281\fdg035 ; ~ -7\fdg353   &   shock-heated region   \\
P2 &    745201101      &   281\fdg227 ; ~ -7\fdg557    &  shock-heated region + stellar     \\
P3  &     745201172    &   280\fdg843 ; ~ -7\fdg256    &   shock-heated region + stellar   \\
P4  &   745201129   & 281\fdg198 ; ~ -7\fdg318        &  shock-heated region + stellar       \\
P5$^{\dagger}$ & 745201145  & 280\fdg952 ; ~ -7\fdg338   &  photoionized region    \\
P6$^{\dagger}$ & 745201148  & 280\fdg967 ; ~ -7\fdg445   &  photoionized region   \\
P8$^{\dagger}$  &   745201038       &  280\fdg490 ; ~ -7\fdg756    & photoionized region + stellar     \\
P9$^{\dagger}$   &   745207037   & 281\fdg399 ; ~ -7\fdg275        &  photoionized region      \\
P10$^{\dagger}$   &   745201142   & 281\fdg076 ; ~ -7\fdg421         &  photoionized region      \\
P12$^{\ddagger}$   &   745201219   & 280\fdg692 ; ~ -8\fdg160        &  photoionized region       \\
P16$^{\ddagger}$   &   745201151   & 280\fdg765 ; ~ -7\fdg271       & photoionized region + stellar  \\
\hline
\end{tabular}
\begin{tablenotes}
\small
\item {$^{\dagger}$ and $^{\ddagger}$  See the text (Section~\ref{sec:gas_conditions}).} 
\item {$^{\ast}$ Based on visual check, followed by an analysis of the [S\,{\sc ii}]/H$\alpha$  ratio from the LAMOST spectra.} 
\end{tablenotes}
\end{threeparttable}
\end{table*}

\begin{table*}
\centering
\caption{Relative line fluxes from the RTT150 spectra taken from different locations with the 1-$\sigma$ errors. The measured [S\,{\sc ii}]/H$\alpha$ and  [N\,{\sc ii}]/H$\alpha$ emission line ratios are also listed.}
\label{tab:Table4b}
 \begin{threeparttable}
 \begin{tabular}{@{}p{3.0cm}p{1.8cm}p{1.8cm}p{1.8cm}p{1.8cm}@{}}
 \hline
                             &  S2     &  S3     &	 NW        &	N 	                     	 \\[0.5 ex]
\hline												
H$\beta$ $\lambda$4861	      &	$20\pm4$ &	$13\pm1$ &	$23\pm3$	&	$19\pm5$				  \\	

[\ion{O}{iii}] $\lambda$4959 &	$9\pm4$	 &	$19\pm3$	&	$25\pm3$	&	$7\pm1$	   		   \\
												
[\ion{O}{iii}] $\lambda$5007 &   $6\pm2$	&	$14\pm1$	&	$22\pm3$	&	$10\pm1$    		  \\
												
[\ion{O}{i}] $\lambda$6300 	&	$26\pm3$ &	$44\pm6$	&	$20\pm3$	&	$37\pm10$					 \\
												
[\ion{O}{i}] $\lambda$6363 	&	$25\pm4$  &	$23\pm6$	 	&	$18\pm3$    &	$57\pm10$	 		 \\
												
[\ion{N}{ii}] $\lambda$6548 &	 $35\pm11$ &	$49\pm14$  	&	$...$	    &	$34\pm15$					  \\
												
H$\alpha$ $\lambda$6563 	 &	$100\pm13$  &	$100\pm16$	  &	$100\pm11$	&	$100\pm16$  			 \\
												
[\ion{N}{ii}] $\lambda$6584  &  $42\pm11$  &	$25\pm9$  &	$24\pm3$	&	$49\pm16$    	 		 \\
												
[\ion{S}{ii}] $\lambda$6716  &	$25\pm6$ &	$26\pm5$	&	$24\pm10$	&	$25\pm12$			    \\
												
[\ion{S}{ii}] $\lambda$6731  &	$42\pm10$  &	$24\pm4$	&	$21\pm8$	&	$28\pm13$  			 \\[0.5 ex]

$F$(H$\alpha$)$^{\dagger}$   &	$3.2\pm0.4$ &	$4.2\pm0.7$   &	$3.3\pm0.4$ &	$2.2\pm0.4$ 	 \\

[\ion{S}{ii}]/H$\alpha$ 	 &	 $0.67\pm0.03$ &	   $0.50\pm0.01$   & $0.45\pm0.15$  &	   $0.53\pm0.15$ 	  \\[0.5 ex]

[\ion{N}{ii}]/H$\alpha$ 	  &	$0.77\pm0.13$ &	  $0.75\pm0.11$ &  $0.24\pm0.01$  &	   $0.83\pm0.18$ 	  \\[0.5 ex]
\hline  
\end{tabular}
\begin{tablenotes}
\item {$^{\dagger}$ The fluxes are presented in units of $10^{-15}$~erg~cm$^{-2}$~s$^{-1}$~\AA$^{-1}$.} 
\end{tablenotes}
\end{threeparttable}
\end{table*}

\clearpage \clearpage
\section{Discussion}
\label{sec:discuss}
\subsection{Optical morphology}
A previous study by \citet{Ga11} reported strong H$\alpha$ filaments in the northern region from SHASSA image \citep{Ga01}; however, the H$\alpha$ emission appears weak within the shell area of the SNR (see their fig.~6, lower panel). Therefore, they considered the H$\alpha$ emission in this direction to be probably not physically associated with the SNR. 

The lack of radio emission in the northern region of the SNR, where H$\alpha$ emission is dominant, can be attributed to the blowout scenario (e.g. \citealt{Dubner2015}). There are also cases where filaments or diffuse H$\alpha$ emission are observed independently, without any clear correlation to other wavelengths (e.g. \citealt{St08, Fe10}). The discrepancy between the distributions of optical and radio emissions could be attributed to the large size of the SNR, its evolutionary stage, and the properties of the local interstellar medium (ISM) where the shock wave interacts (e.g. VRO~42.05.01: \citealt{Pi87}, HB3: \citealt{Fe95}, W44: \citealt{Gi97}, and G159.6+7.3: \citealt{Fe10}). 

Our continuum-subtracted H$\alpha$ images clearly reveal a filamentary morphology in the S1, S2, and S3 regions. In contrast, the N region displays a faint circular structure, while the NW region shows predominantly diffuse emission (see Fig.~\ref{fig:figure2b}).

\subsection{Optical spectral line ratios and gas conditions}
\label{sec:gas_conditions}
To investigate its physical properties and environment, we study G25.1$-$2.3 using LAMOST and RTT150 spectroscopy at various locations. The spectra obtained from many regions reveal distinct emission line properties, as shown in Tables~\ref{tab:Table3} and \ref{tab:Table4b}. We measured the ratios of several emission lines to determine the physical parameters of the gas. 

\subsubsection{Physical properties of the SNR from LAMOST spectra}
Optical emission arising from shock-excited interstellar gas, rather than from photoionized gas, can be recognized by spectra showing a line ratio of [S\,{\sc ii}]/H$\alpha$  $\geq$  0.4 (e.g. \citealt{Blair1981, Dopita1984, Fe85, Long2017}). In addition, the [N\,{\sc ii}]/H$\alpha$ line ratio (typically in the range 0.5$-$1.5, see \citealt{Fe85}) is also widely used as a diagnostic to distinguish shock-heated gas from photoionized regions.

For the P1, P2, P3, and P4 positions, the [S\,{\sc ii}]/H$\alpha$ ratios ranging from approximately 0.51 to 0.83 indicate that the emission is likely associated with shock-excited regions. The [N\,{\sc ii}]/H$\alpha$ line ratios also support these results ($\sim$0.52$-$1.35), except for the position P3 (see Table~\ref{tab:Table3}).  Another diagnostic approach based on optical emission-line ratios, incorporating both log(H$\alpha$/[S\,{\sc ii}]) and log(H$\alpha$/[N\,{\sc ii}]), has been applied by several authors (e.g. \citealt{FrewParker2010, Sabin13}). The average values derived from the LAMOST, log(H$\alpha$/[S\,{\sc ii}]) ranging from $\sim$0.08 to
0.29 and log(H$\alpha$/[N\,{\sc ii}]) from $-0.13$ to 0.28, also place these positions within the region typically occupied by SNRs (see the top panel of fig. 7 in \citealt{Sabin13}). When the measured values for the P1–P4 regions are plotted on the diagnostic diagram of log(H$\alpha$/[S\,{\sc ii}]) versus the  [S\,{\sc ii}] $\lambda$6716/$\lambda$6731 ratio (bottom panel of fig. 7 in \citealt{Sabin13}), they also fall within the area associated with SNRs.

For the P5, P6, P8, P9, and P10 positions, the calculated [S\,{\sc ii}]/H$\alpha$ ratios ($\sim$0.16$-$0.29) and [N\,{\sc ii}]/H$\alpha$ ratios ($\sim$0.23$-$0.44) suggest that the emission originates from photoionized regions. For these positions, the mean values of log(H$\alpha$/[S\,{\sc ii}]), ranging from $\sim$0.54 to 0.80, and log(H$\alpha$/[N\,{\sc ii}]), spanning $\sim$0.36 to 0.64, further place these positions within the parameter space typically associated with photoionized regions (see the top panel of fig. 7 in \citealt{Sabin13}).

In the P12 and P16 spectra, [S\,{\sc ii}] $\lambda$$\lambda$6716, 6731 lines cannot be determined reliably. However, on the basis of the average \texttt {rms} values of these spectra at these wavelengths, an upper value for the fluxes of the [S\,{\sc ii}] emission lines can be calculated. For these two spectra, the [S\,{\sc ii}]/H$\alpha$ values calculated from the [S\,{\sc ii}] line fluxes estimated by this method remain below the average 0.4 value. This strengthens the possibility that these are photoionized regions. Similarly, the flux of the [S\,{\sc ii}] $\lambda$6731 line could not be measured at positions P5, P6, P8, P9 and P10. If we consider the same idea, the fluxes of the [S\,{\sc ii}] $\lambda$6731 line estimated from the average \texttt {rms} values can be included in the [S\,{\sc ii}]/H$\alpha$ ratio. In this case, the value of [S\,{\sc ii}]/H$\alpha$ does not exceed 0.4. Therefore, it would be appropriate to classify these regions as photoionized regions.

The electron density can be found using the [S\,{\sc ii}] $\lambda$6716/$\lambda$6731 ratio, which is available from spectral data \citep{OsFe06}. The LAMOST spectra at the P1, P2, P3, and P4 positions allowed us to estimate the electron density (see Table~\ref{tab:Table5}). The [S\,{\sc ii}] $\lambda$6716/$\lambda$6731 line ratios for P1 and P3 are close to the low density limit of $\sim$1.4, indicating the electron density of $\sim$120~cm$^{-3}$. For position P2, this ratio was observed to be around 1.02, indicating a higher density ($\sim$640 cm$^{-3}$) compared to other locations (P1, P3, and P4). This is a relatively high value and implies a correspondingly high pre-shock cloud density.

For the P1, P2, P3, and P4 positions, we chose a shock velocity of $V_{\rm s} =100$ km~s$^{-1}$ and estimated pre-shock densities ($n_{\rm c}$) using the formula given by \citet{Dopita1979}. We found that $n_{\rm c}$ ranges from 2.7 to 14.2 cm$^{-3}$. The upper end of this range (for the P2 position) suggests that the optical emission may result from the SNR's shock interaction with dense interstellar clouds.

\begin{table}
\centering
 \caption{Electron density ($n_{\rm e}$) derived from the observed [S\,{\sc ii}] $\lambda$6716/$\lambda$6731 line ratio obtained from the LAMOST and RTT150 spectra, assuming an electron temperature of $T=10^{4}$~K.}
 \label{tab:Table5}
 \begin{tabular}{@{}p{1.3cm}p{2.7cm}p{2.3cm}@{}}
 \hline
Slit ID & [S\,{\sc ii}] $\lambda$6716/$\lambda$6731 & Electron density   \\
 &  & (cm$^{-3}$)  \\
\hline
P1  &  $1.35\pm0.11$ & $120\pm10$    \\
P2  &   $1.02\pm0.02$  & $640\pm70$   \\
P3 &   $1.36\pm0.10$  & $120\pm10$   \\
P4 &   $1.24\pm0.06$  & $250\pm40$   \\
\hline
S2  &  $0.58\pm0.01$ & $4500\pm500$     \\
S3 &   $1.08\pm0.01$  & $490\pm30$   \\
NW &   $1.17\pm0.01$  & $390\pm20$   \\
N &   $0.88\pm0.01$  & $1030\pm10$   \\

\hline
\end{tabular}
\end{table}

\subsubsection{Physical properties of the SNR from RTT150 spectra}
We measured [\ion{S}{ii}]/H$\alpha$ ratios ranging from 0.45 to 0.67 for the S2, S3, NW, and N regions. These values indicate that the observed emission arises from shock-heated gas. With the exception of region NW, the [N\,{\sc ii}]/H$\alpha$ ratios ($\sim$0.75$-$0.83) further support this interpretation (see Table~\ref{tab:Table4b}). The corresponding logarithmic ratios are log(H$\alpha$/[\ion{S}{ii}])  $\sim$ 0.17$-$0.35 and log(H$\alpha$/[N\,{\sc ii}]) $\sim$ 0.08$-$0.13. According to the diagnostic diagrams presented by \citet{Sabin13}, these values fall within the region characteristic of SNRs. Moreover, considering that the [\ion{S}{ii}] $\lambda$6716/$\lambda$6731 ratios lie between 0.58 and 1.17, our measurements also fall within the SNR zone in the log(H$\alpha$/[\ion{S}{ii}]) versus [\ion{S}{ii}] $\lambda$6716/$\lambda$6731 diagnostic diagram presented by \citet{Sabin13}.

\begin{figure*}
\includegraphics[angle=0, width=7.9cm]{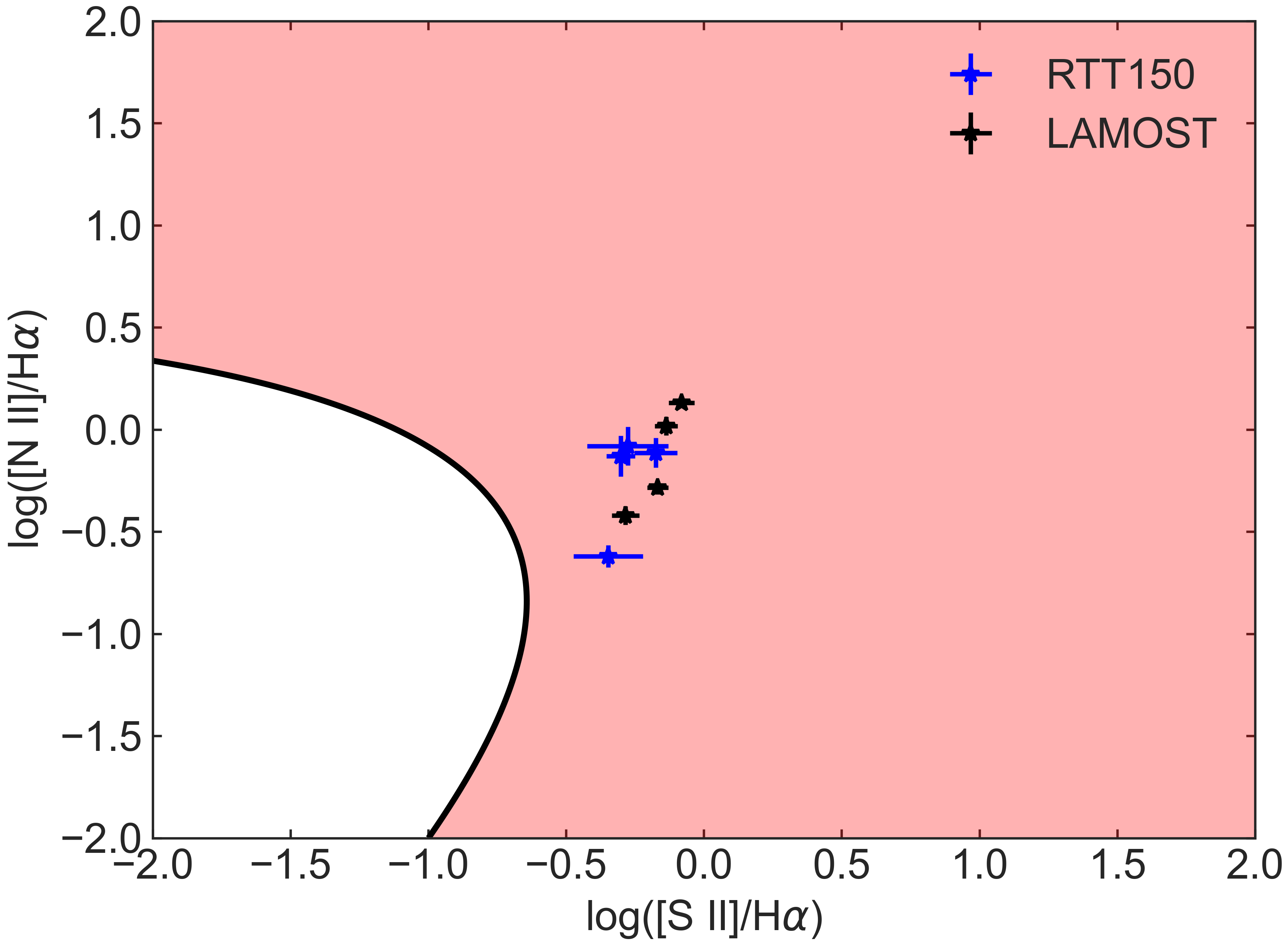}
\includegraphics[angle=0, width=7.9cm]{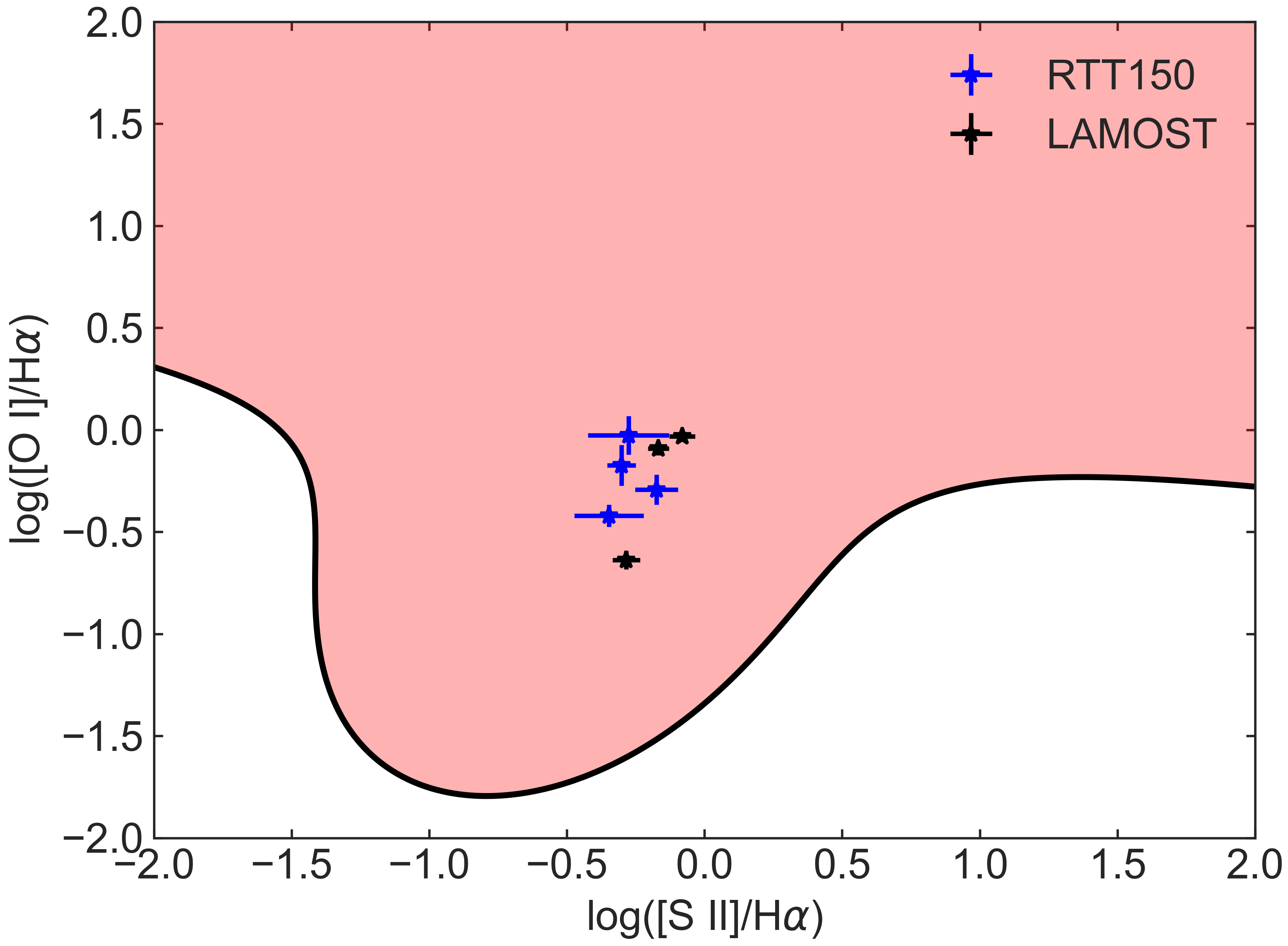}
\includegraphics[angle=0, width=7.9cm]{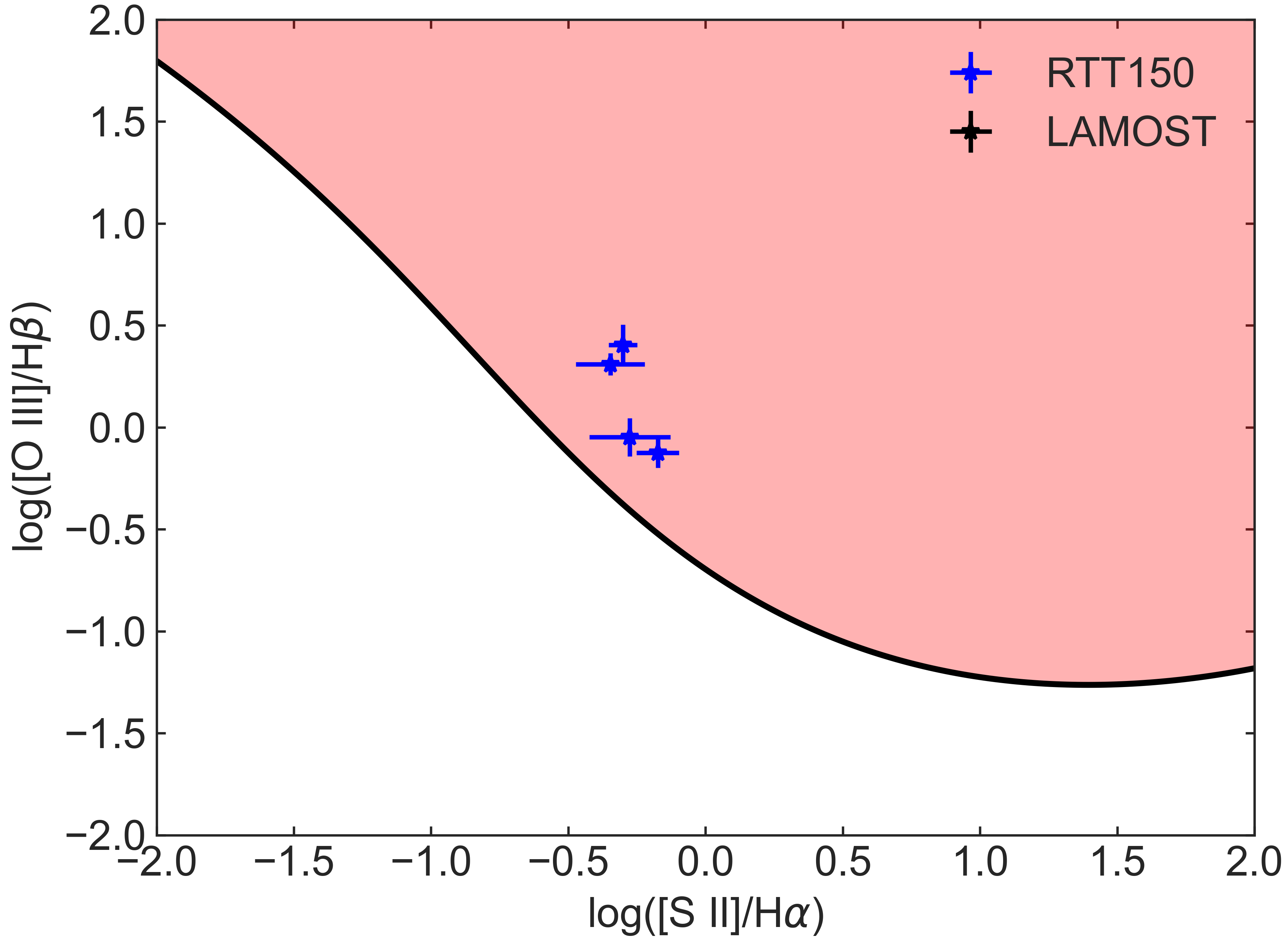}
\includegraphics[angle=0, width=7.9cm]{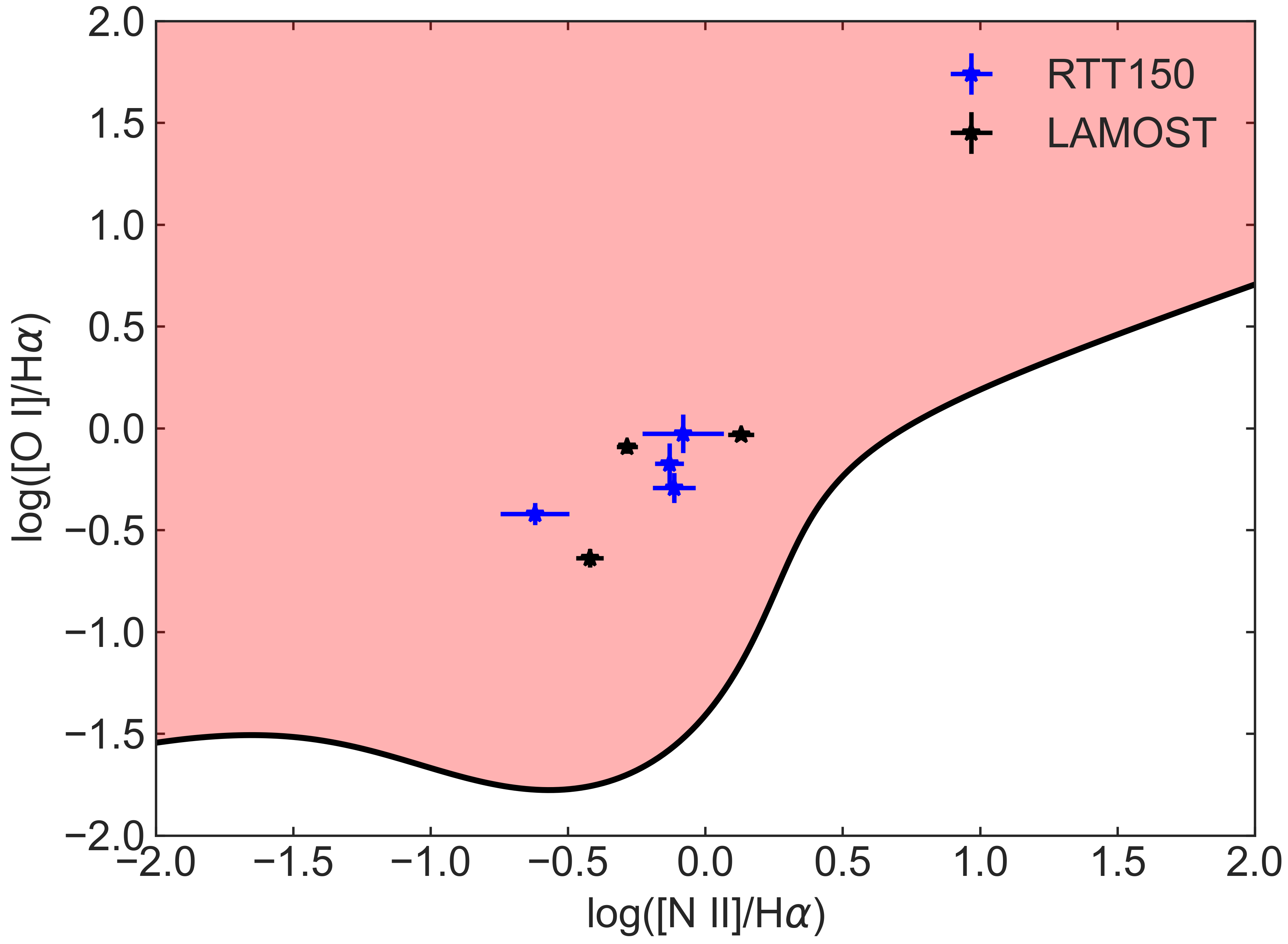}
\includegraphics[angle=0, width=7.9cm]{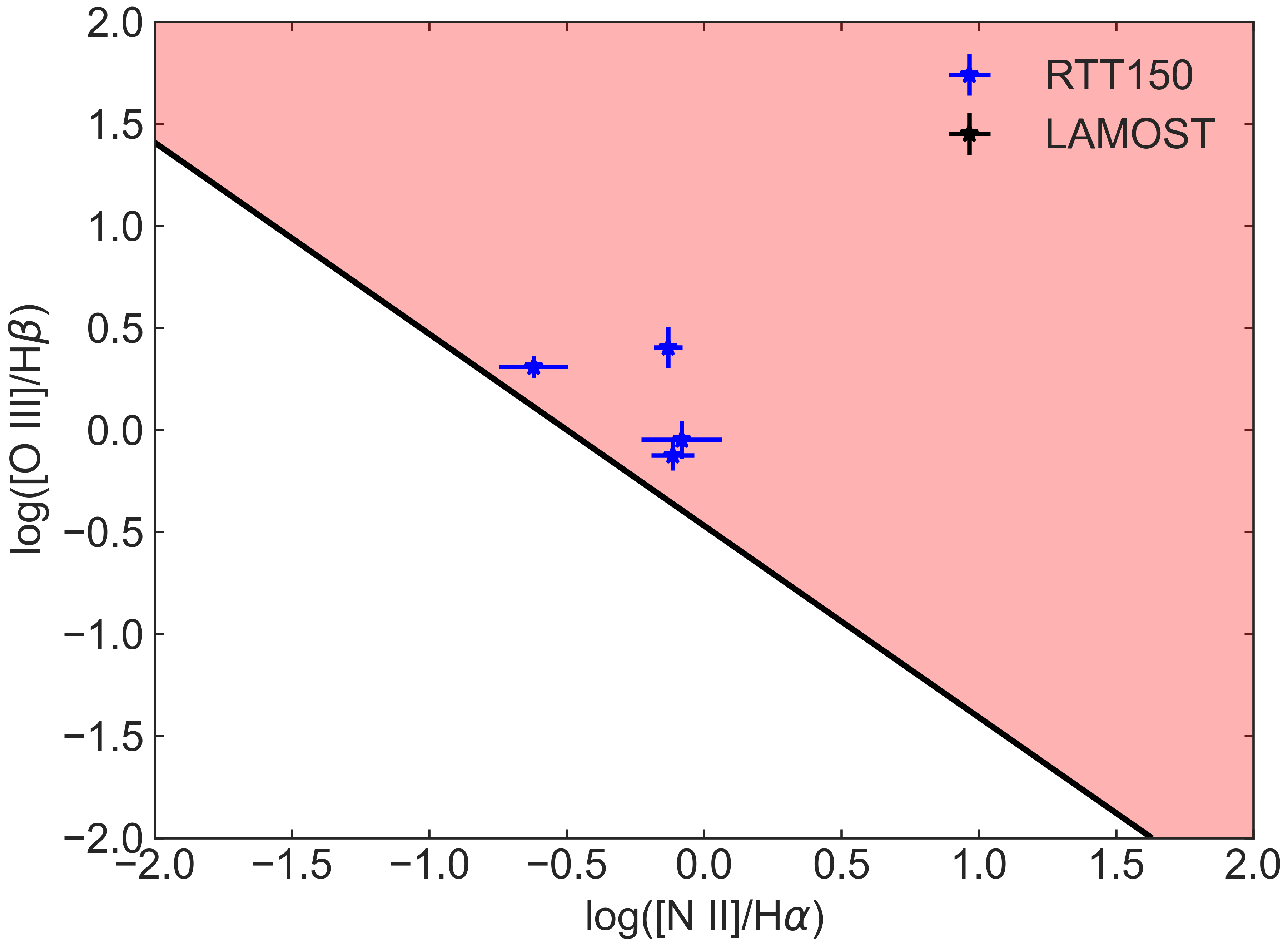}
\includegraphics[angle=0, width=7.9cm]{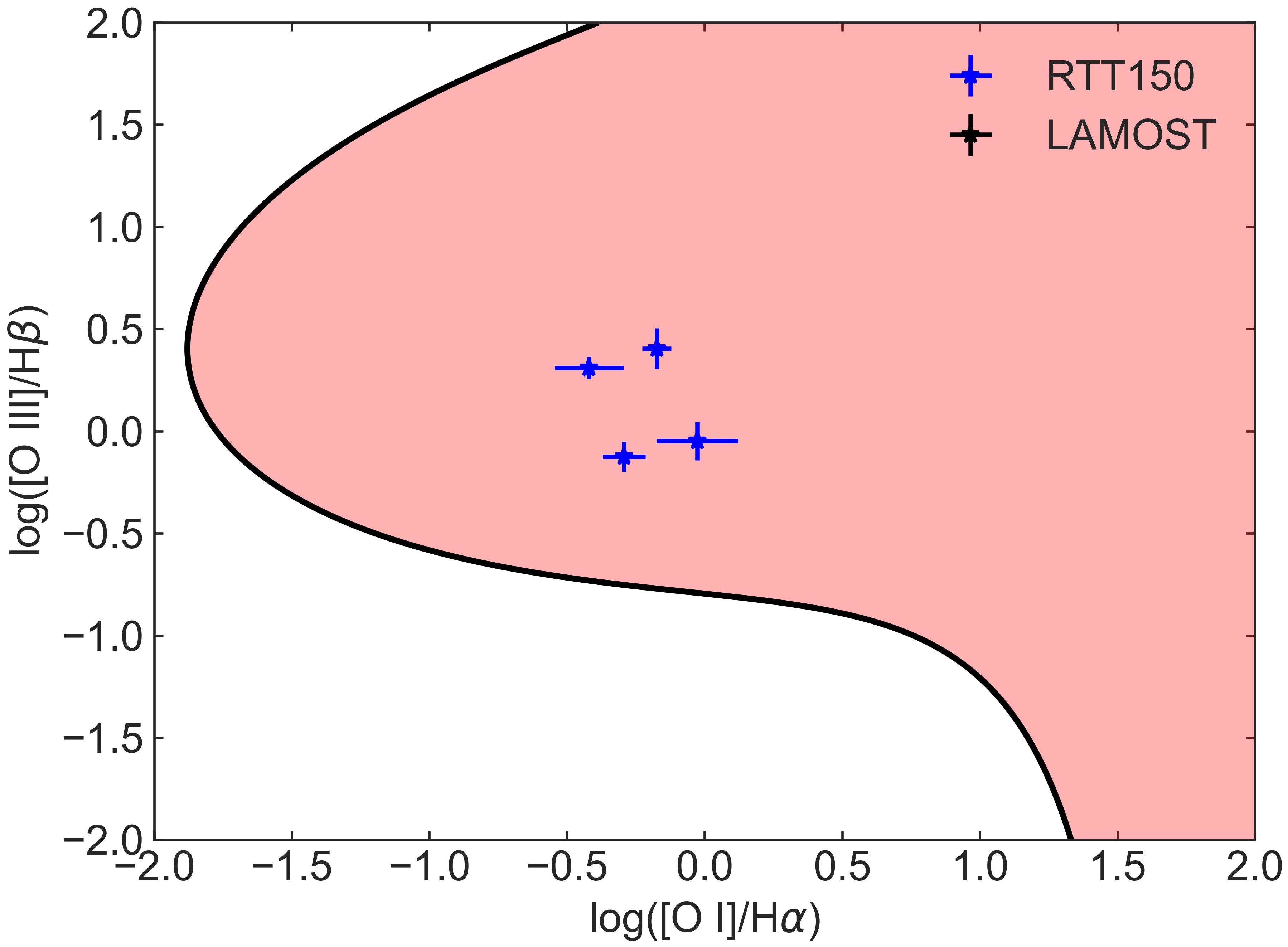}
\caption{Two-dimensional diagnostic diagrams developed by \citet{Kopsacheili2020}, overlaid with our spectroscopic measurements from RTT150 (blue asterisks) and LAMOST (black asterisks). The pink shaded region indicates the parameter space characteristic of shock-excited gas. The diagnostic datasets were accessed through the link provided by  \citet{Kopsacheili2024}.}
\label{fig:diagrams}
\end{figure*}

To further investigate the shock-heated nature of the S2, S3, NW, and N regions, we present diagnostic diagrams in Fig.~\ref{fig:diagrams}, adapted from \citet{Kopsacheili2020}, which distinguish between shock-heated and photoionized regions using various emission-line ratios. We also include the measured ratios for P1$-$P4 from the LAMOST data. The line ratios derived from both the RTT150 and LAMOST datasets meet the required criteria and fall within the parameter space characteristic of shock-heated regions.

The electron density ($n_{\rm e}$) was estimated to range between approximately 390 and 4500 cm$^{-3}$, using the [\ion{S}{ii}] $\lambda$6716/$\lambda$6731 doublet line ratio (see Table~\ref{tab:Table5}). The spectra from the S2 region indicate a highly dense medium ($n_{\rm e}$ $\sim$ 4500 cm$^{-3}$). Considering all the values presented in Table~\ref{tab:Table5}, which correspond to the southern and northern regions of the SNR, significant variations in electron density are evident. These variations are indicative of intrinsic fluctuations within the medium (e.g. IC 443:  \citealt{Ba24, Alarie2019}). Another noteworthy point in Table~\ref{tab:Table5} is the discrepancy between the electron densities ($n_{\rm e}$) derived from the LAMOST and RTT150 spectra, particularly for regions P4 ($\sim$250~cm$^{-3}$) and N ($\sim$1030~cm$^{-3}$), which are located close to each other (see Fig.~\ref{fig:figure2a}). Despite their proximity, the electron density inferred for N is roughly four times higher than that for P4. The differences in the derived line ratios and electron densities arise primarily from instrumental resolution and intrinsic ISM structure. The higher resolution of LAMOST (R $\sim$ 7500) enables a cleaner separation of blended lines such as the [\ion{S}{ii}] $\lambda$$\lambda$6716, 6731 doublet, while the lower-resolution RTT150 (R $\sim$ 749) spectra introduce larger uncertainties in flux measurements. In addition, small-scale inhomogeneities in the ISM can produce noticeable variations even between nearby positions, as noted in previous studies (e.g. \citealt{Domcek2023}). These combined effects naturally account for the observed discrepancies between the datasets obtained from the two nearby regions. 

We also measured [\ion{O}{iii}] $\lambda$5007/H$\beta$ ratios ranging from 0.3 to 1.1. When expressed in logarithmic form ($-0.52$ to 0.04) and compared with the diagnostic diagrams of \citet{Sabin13} in their fig. 8, these values fall in the region populated by known SNRs. This holds true in both log([\ion{O}{iii}] $\lambda$5007/H$\beta$) versus log([\ion{S}{ii}]/H$\alpha$) and log([\ion{O}{iii}] $\lambda$5007/H$\beta$) versus log([\ion{N}{ii}] $\lambda$6584/H$\alpha$) diagrams. Therefore, the behavior of the [\ion{O}{iii}] $\lambda$5007/H$\beta$ ratio provides additional confirmation that the observed emission is produced by shock-heated gas typical of SNRs.

Based on the [\ion{O}{iii}] ($\lambda$4959+$\lambda$5007)/H$\beta$ ratios and adopting  the shock model of \citet{Hartigan1987}, the shock velocity ($V_{\rm s}$) was estimated to be approximately 100 km~s$^{-1}$. Combining the derived electron densities with this velocity, we estimated the corresponding pre-shock densities ($n_{\rm c}$) to range between 9 and 100 cm$^{-3}$, following the relation of \citet{Dopita1979}. 

The logarithmic extinction coefficient ($c$) was derived from the observed H$\alpha$/H$\beta$ ratios, yielding values between 0.49 and 1.24, based on the relation given by \citet{Kaler1976}. Using the relation $E(B-V) = 0.77c$ \citep{OsFe06}, the corresponding color excess values were found to be approximately 0.38$-$0.96, which correspond to visual extinctions of $A_{\rm V}$ $\sim$ 1.18$-$2.98. The hydrogen column density, $N_{\rm H}$, was subsequently estimated to be (2.1$-$5.2) $\times$ $10^{21}$ cm$^{-2}$ based on the relation given by \citet{Predehl1995}.

Spectral analysis across the regions reveals significant variations in reddening, electron density, and pre-shock density, consistent with multiple dust clouds, large-scale shock expansion asymmetries, and density inhomogeneities in the surrounding medium. Similar variations have been reported for other Galactic SNRs, including CTB~1 \citep{Fesen1997}, G190.9$-$2.2 \citep{Bakis2025b}, G203.1+6.6 and G152.4$-$2.1 \citep{Aktekin2025}.

\subsection{Evolutionary status of G25.1$-$2.3}
 
The observed optical emission aligns with radiative shocks, which may result from SNRs that have progressed beyond the adiabatic phase (e.g. Diprotodon: \citealt{Filipovic2024}) or from the interaction of the forward shock with a dense medium (e.g. G7.7$-$3.7: \citealt{Domcek2023}).
  
To determine the evolutionary status of G25.1$-$2.3, we applied the $\Sigma-D$ relation method as described by \citet{Urosevic2020, Urosevic2022}. Using this approach, the evolutionary stages of numerous Galactic (e.g. \citealt{Filipovic2023}) and extragalactic radio SNRs (e.g. \citealt{Ghavam2024, Alsaberi2024}) have been successfully determined.

Based on the $\Sigma$–$D$ relation (\citealt{Pavlovic2018}; their fig. 3), we estimated the position of the remnant’s radio surface brightness ($\Sigma_{\rm {1GHz}}\sim5.0\times10^{-22}$ W m$^{-2}$ Hz$^{-1}$ sr$^{-1}$ from \citealt{Ga11}) and diameter ($D$=31$-$35 pc from \citealt{Urosevic2025}) on the $\Sigma$–$D$ plane. Our result suggests that the SNR is in the late Sedov phase of its evolution, with an average explosion energy of approximately 0.5 $\times$ 10$^{51}$ erg, expanding into an environment characterized by a moderate to relatively high ambient density of $0.2-0.5$~cm$^{-3}$. This density range is not consistent with the much higher values derived from our spectroscopic analysis. This discrepancy likely arises because $\Sigma$–$D$ relations reflect large-scale, average ISM conditions, whereas spectroscopy probes locally enhanced densities. Although theoretical $\Sigma$–$D$ relation assume a homogeneous ambient medium, typically modeled with densities ranging from 0.005 to 2~cm$^{-3}$, the actual SNR environment is generally inhomogeneous on the spatial scales of the remnant (see \citealt{Kostic2016}). The observed optical filaments are most likely produced by the interaction of the SNR with dense ambient material in the observed regions.

Additionally, we determined the SNR's magnetic field strength using the equipartition (eqp) method \citep{Arbutina2012, Arbutina2013, Urosevic2018}. 

For spectral indices of $\alpha$ = $-0.49$ and $-0.50$, we applied the adopted eqp calculation presented by \citet{Filipovic2023}. This approach has been successfully used to estimate magnetic field strengths in SNRs with $\alpha$ $>$ $-0.5$ (e.g. Ancora: \citealt{Filipovic2023} and Veliki SNR: \citealt{Smeaton2025}). Using the adapted eqp method in \citet{Filipovic2023}, specifically their Equations 7 and 9, we derived magnetic field strengths of $B$ $\sim$ $23$ $\mu$G under the electron equipartition assumption and $B$ $\sim$ $27$ $\mu$G  under proton equipartition for $\alpha$ = $-0.49$. For $\alpha$ = $-0.50$, the corresponding values are 
$B$ $\sim$ $24$ and 
$B$ $\sim$ $28$ $\mu$G, respectively. The shock velocity, $V_{\rm s} = 100~{\rm km~s^{-1}}$, derived from the [\ion{O}{iii}]/H$\beta$ ratio in our optical spectroscopy, 
was used to constrain the physical conditions of the shock and to estimate the 
proton-to-electron energy ratio $\kappa$ under the assumption of a strong, 
non-modified shock and adopting $T_{\rm e} = 0.1\,T_{\rm p}$. The shock velocity provides an important constraint on the shock regime and associated
equipartition parameters. For the magnetic field calculations, we adopted the following parameters: $S_{\rm 1\,GHz}$=8~Jy, $d$=2.3~kpc (within the 2.2$-$2.5~kpc range reported by \citealt{Urosevic2025}), angular radius $\theta$=27.5~arcmin, and filling factor $f$=0.25.

For $\alpha$ = $-0.51$, using the online calculator\footnote{\url{https://poincare.matf.bg.ac.rs/~arbo/eqp/}} and assuming $\kappa$=0 (see \citealt{Urosevic2018}), we estimated a magnetic field strength of $B$ $\sim$ 17.3 $\mu$G and a corresponding minimum energy of $E_{\rm min}$ = 5.3$\times10^{48}$~erg.

To draw a conclusion about the evolutionary stage of G25.1$-$2.3, we consider the combined results from the $\Sigma$–$D$ and the eqp methods (see \citealt{Urosevic2022}). The $\Sigma$–$D$ track suggests that the SNR in the late Sedov phase, expanding in a moderate to relatively high ambient density of $0.2-0.5$~cm$^{-3}$, with an average explosion energy of $\sim$0.5 $\times$ 10$^{51}$ erg. The eqp magnetic field strength is $\sim$20~$\mu$G. For an SNR expanding in such an ambient density, a magnetic field of around 20 $\mu$G can be produced by compression of the ISM magnetic field, therefore, magnetic field amplification (MFA) by strong shocks is not necessary. Additionally, a spectral index of about $-0.5$ suggests that the SNR is not in a young evolutionary stage. Taken together, these characteristics indicate that G25.1$-$2.3 is an evolved SNR in the late Sedov phase.

\subsection{H\,{\sc i} and CO environment}
\label{sec:hi_co_environment}

\begin{figure*}
\includegraphics[angle=0, width=\linewidth]{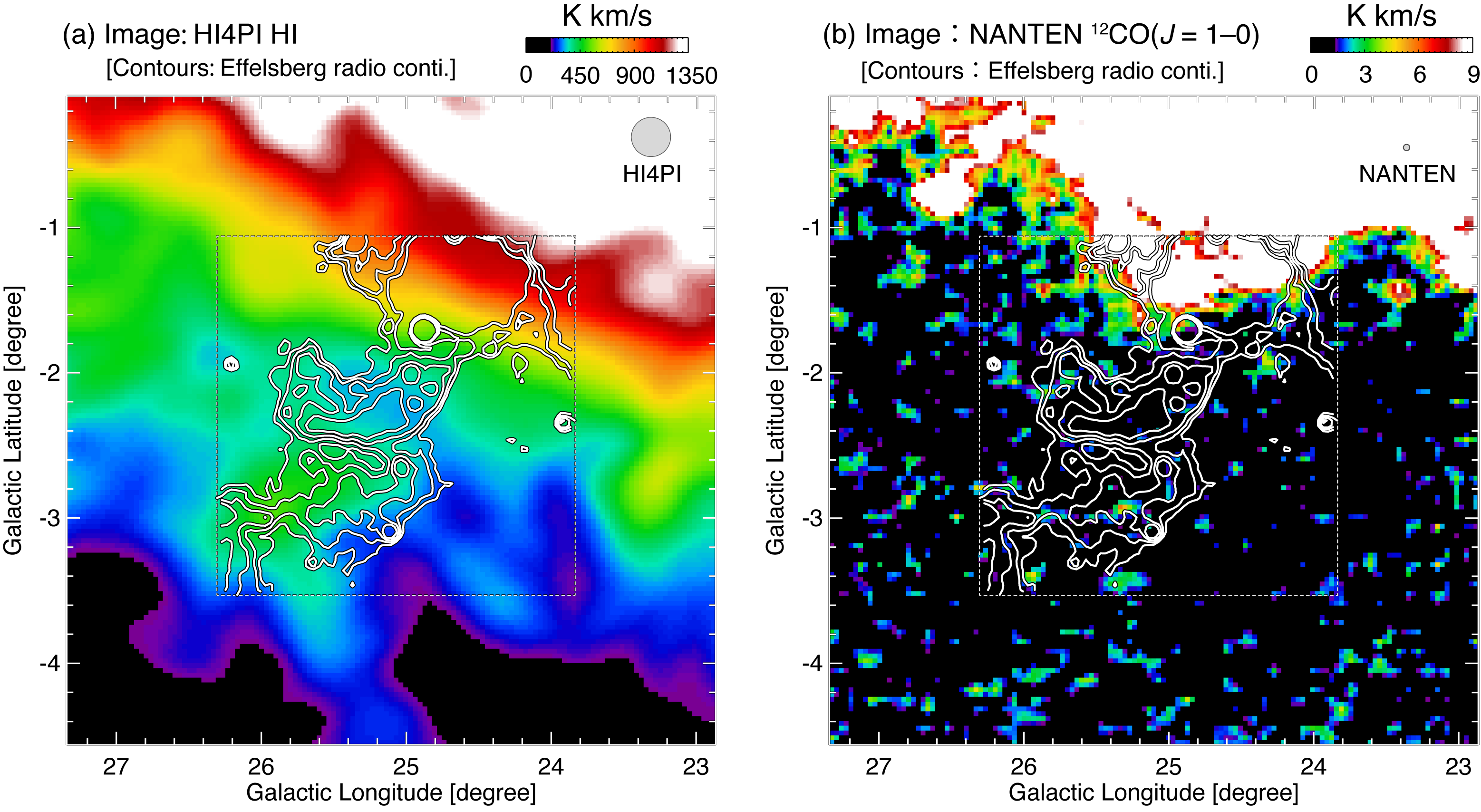}
\caption{Integrated intensity maps of (a) HI4PI H\,{\sc i} \citep{HI4PI} and (b) NANTEN $^{12}$CO($J$~=~1--0) \citep{NANTEN}. The integration velocity range of H{\sc i} and CO is 40--58 km~s$^{-1}$. Superposed contours represent the 2695~MHz radio continuum obtained from Effelsberg \citep{Effelsberg}. The contour levels are 100, 130, 160, 190, 220, and 250 K.  The white--dashed line and red-solid line show the observed areas of the radio continuum.}
\label{fig:co_hi_integ}
\end{figure*}

In order to explore the interstellar gaseous environment surrounding the SNR G25.1$-$2.3, we carried out H\,{\sc i} and CO analysis. Fig.~\ref{fig:co_hi_integ} shows the velocity-integrated intensity maps of H\,{\sc i} and $^{12}$CO($J$~=~1--0) obtained from the HI4PI survey \citep{HI4PI} and the NANTEN CO survey \citep{NANTEN}. The angular resolutions of H\,{\sc i} and CO are $\sim$$16'$ and $\sim$2\farcm6, respectively. We found no dense molecular cloud in CO, while a cavity-like distribution of H\,{\sc i} is seen toward the radio continuum shell. Although the velocity range $V_{\mathrm{LSR}}$: 40--58 km s$^{-1}$ is slightly different from the previous study \citep[$V_{\mathrm{LSR}}$: 37.1--41.2~km~s$^{-1}$,][]{Ga11}, the general trend is similar to each other.

\begin{figure*}
\includegraphics[angle=0, width=10cm]{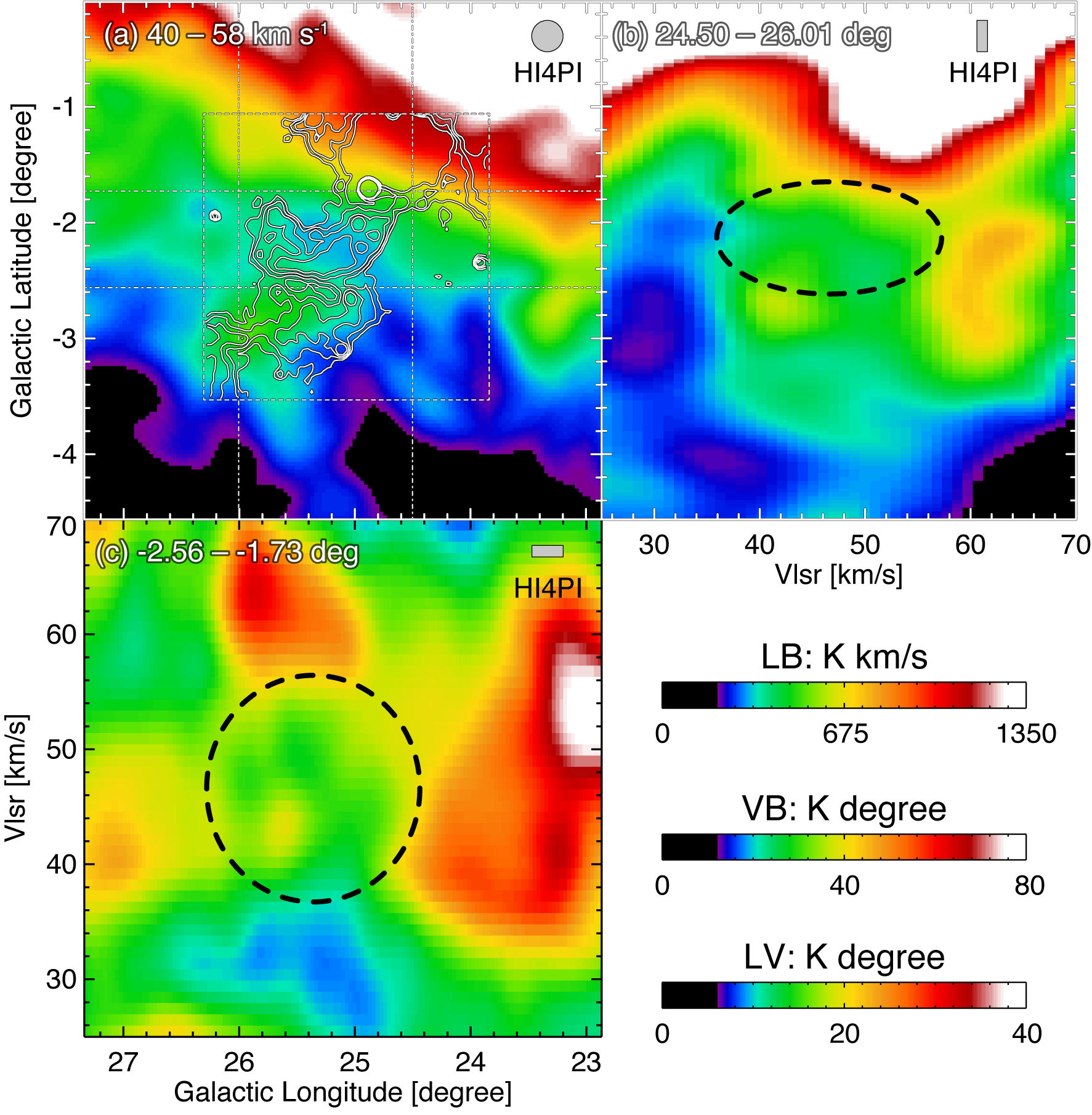}
\caption{(a) Integrated intensity map of H\,{\sc i}. The integration velocity range and the overlaid contours are the same as those shown in Fig.~\ref{fig:co_hi_integ}a. The dashed--dotted lines show the integration ranges in Galactic latitude and longitude. (b) Galactic latitude--velocity diagram of H\,{\sc i}. The integration range in latitude is from $-2.56$ to $-1.73$. (c) Galactic longitude--velocity diagram of H\,{\sc i}. The integration range in longitude is from 24\fdg50 to 26\fdg01. The dashed circles in panels (b) and (c) represent an expanding gas motion (see the text).}
\label{fig:hi_pv}
\end{figure*}

Fig.~\ref{fig:hi_pv} shows the position--velocity diagrams of H\,{\sc i}. We newly found a hole-like distribution of H\,{\sc i} for each p--v diagram, whose spatial extent is roughly consistent with the diameter of the SNR. This indicates that the hole-like H\,{\sc i} distributions in p--v diagrams represent the expanding gas motion formed by strong progenitor wind and/or supernova shocks \citep[e.g.,][]{Koo90,Koo91}. In that case, the expanding velocity of H\,{\sc i} is to be $\sim$10~km~s$^{-1}$. Although this cannot rule out the association of the H\,{\sc i} cloud/hole at $V_{\mathrm{LSR}}$: 37.1--41.2~km~s$^{-1}$ that was previously mentioned, the H\,{\sc i} hole possibly appeared due to the H\,{\sc i} absorption line behind the bright radio continuum of the SNR.

\section{Conclusions}
\label{sec:conc}
We have presented optical imaging and spectroscopic analysis of the SNR G25.1$-$2.3, and also analysed the H\,{\sc i} and CO radio data to investigate its environment.

H$\alpha$ images of the northern and southern regions exhibit filamentary and diffuse structures. Using LAMOST and RTT150 data, some prominent emission lines, such as H$\alpha$ $\lambda$6563, [N\,{\sc ii}] $\lambda$6584, and [S\,{\sc ii}] $\lambda$$\lambda$6716, 6731, are detected in many regions. The LAMOST spectra of north region (P1$-$P4 positions) revealing [S\,{\sc ii}]/H$\alpha$ ratios ranging from approximately 0.51 to 0.83. The RTT150 spectra at the S2, S3, NW, and N positions show [S\,{\sc ii}]/H$\alpha$ ratios ranging from 0.45 to 0.67.
Together with other diagnostic line ratios, such as [\ion{O}{i}] $\lambda$6300/H$\alpha$, [N\,{\sc ii}]/H$\alpha$, and [\ion{O}{iii}] $\lambda$5007/H$\beta$, these measurements are fully consistent with standard diagnostic diagrams distinguishing shocks from photoionization. These results provide clear evidence that the optical emission in these regions is dominated by shock-heated gas. The [S\,{\sc ii}] $\lambda$6716/$\lambda$6731 ratio observed in the northern region indicates an electron density of $\sim$$120-1030$ cm$^{-3}$, while in the southern region electron densities ranging from 490 to 4500 cm$^{-3}$. The observed spatial variations in pre-shock density and extinction further suggest that the ambient medium is highly inhomogeneous. A hole-like distribution of H\,{\sc i} was found, with a spatial extent roughly consistent with the diameter of the SNR. To assess the evolutionary stage of G25.1$-$2.3, we employed the $\Sigma$–$D$ relation and estimated its magnetic field strength using the equipartition approach. Together, these diagnostics provide a consistent indication that G25.1$-$2.3 is an evolved remnant in the late Sedov phase.

We found no counterparts in publicly available X-ray or gamma-ray surveys. In particular, future observations with high-resolution X-ray instruments will be essential to further constrain the nature of G25.1$-$2.3.

\section*{Acknowledgements}
We would like to acknowledge the referees, Prof. Dejan {Uro{\v{s}}evi{\'c}} and Milica {An{\dj}eli{\'c}}, for their valuable comments and suggestions that have significantly improved the paper. We also thank Bojan Arbutina for providing useful information on the equipartition calculation. This research was supported by the Scientific and Technological Research Council of T\"{u}rkiye (T\"{U}B\.{I}TAK) through project number 124F089. In this study, observational data obtained within the scope of the project numbered 25ATUG100-3004 and 25ARTT150-3015, conducted using the T100 and RTT150 Telescopes and the TFOSC system at the TUG (T\"{U}B\.{I}TAK National Observatory, Antalya) site under the T\"{u}rkiye National Observatories, have been utilized, and we express our gratitude for the invaluable support provided by the T\"{u}rkiye National Observatories and their personnel. This work has made use of data products from the Guoshoujing Telescope (the Large Sky Area Multi-Object Fiber Spectroscopic Telescope, LAMOST). LAMOST is a National Major Scientific Project built by the Chinese Academy of Sciences. Funding for the project has been provided by the National Development and Reform Commission. LAMOST is operated and managed by the National Astronomical Observatories, Chinese Academy of Sciences. This work has used the image obtained by the Southern H-Alpha Sky Survey Atlas (SHASSA), which is supported by the National Science Foundation. This work was supported by JSPS KAKENHI grant No. 21H01136 (HS), 24H00246 (HS). The NANTEN project is based on a mutual agreement between Nagoya University and the Carnegie Institution of Washington (CIW). We greatly appreciate the hospitality of all the staff members of the Las Campanas Observatory of CIW. We are thankful to many Japanese public donors and companies who contributed to the realization of the project.

\section*{DATA AVAILABILITY}
The T100 and RTT150 data underlying this paper will be shared upon reasonable request to the corresponding author. The LAMOST data are available at https://www.lamost.org/dr9/v2.0/. 

 

\clearpage \clearpage

\bibliographystyle{mnras}
\bibliography{example} 



\appendix
\section{ADDITIONAL SPECTRA}
\begin{figure*}
\includegraphics[angle=0, width=8.2cm]{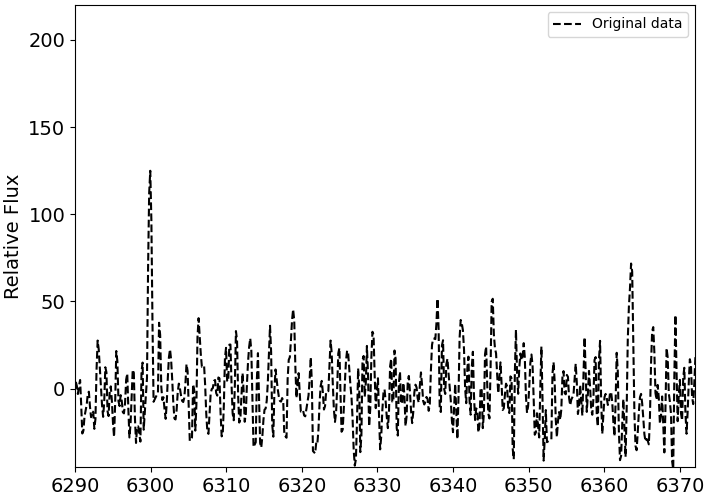}
\includegraphics[angle=0, width=8.2cm]{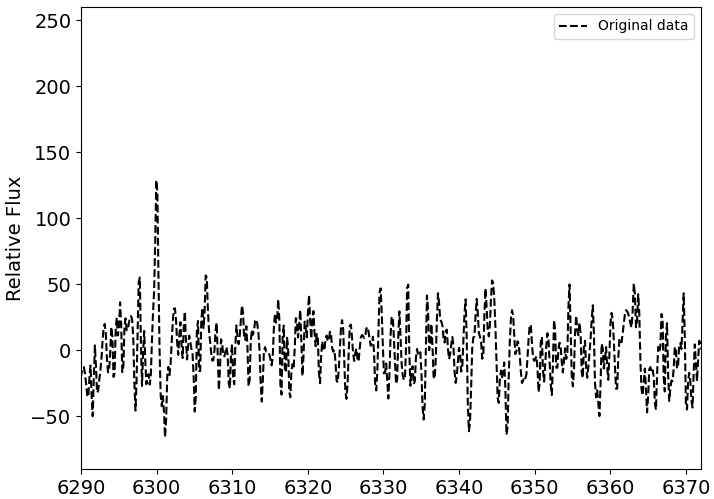}
\includegraphics[angle=0, width=8.2cm]{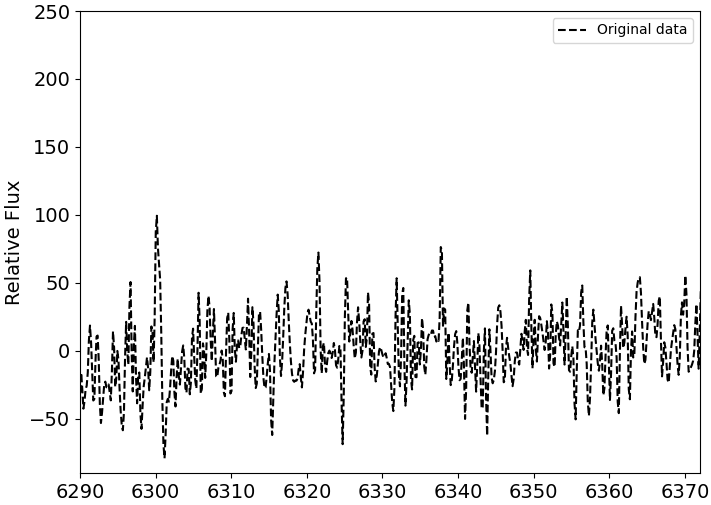}
\includegraphics[angle=0, width=8.2cm]{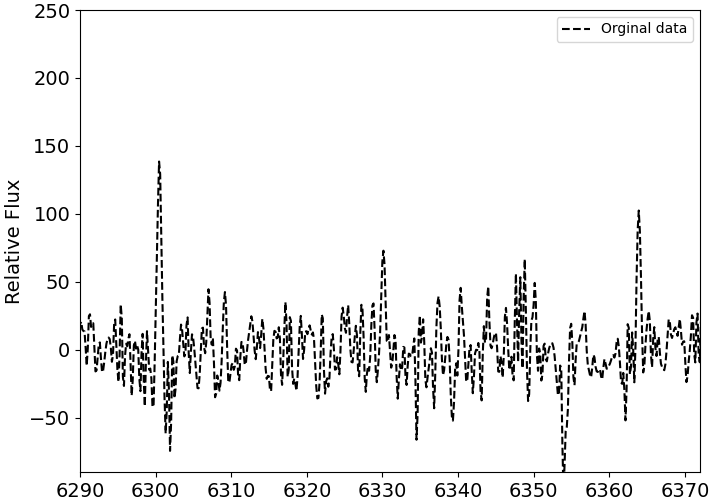}
\includegraphics[angle=0, width=8.2cm]{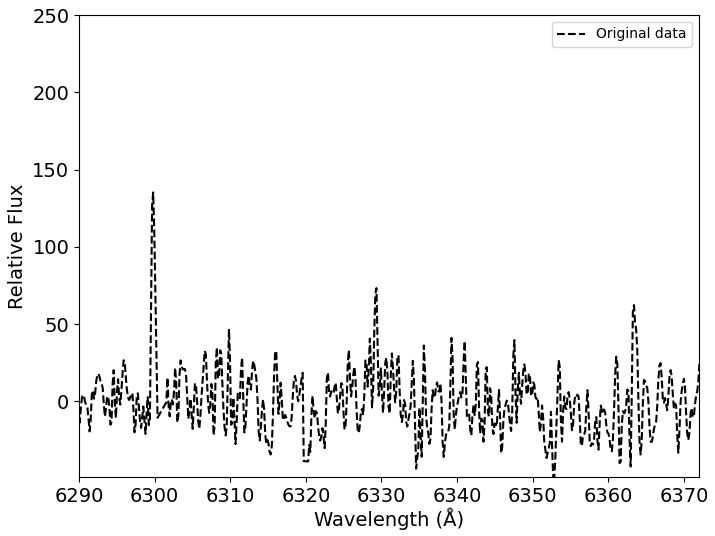}
\includegraphics[angle=0, width=8.2cm]{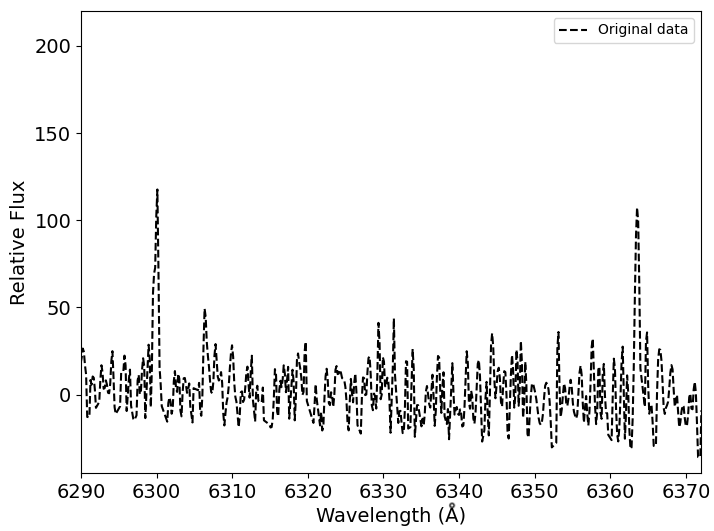}
\caption{LAMOST spectra (6290$-$6370 {\AA}) for P1, P2, P3, P4, P5, and P6 positions.}
\label{FigA1}
\end{figure*}

\begin{figure*}
\includegraphics[angle=0, width=7.5cm]{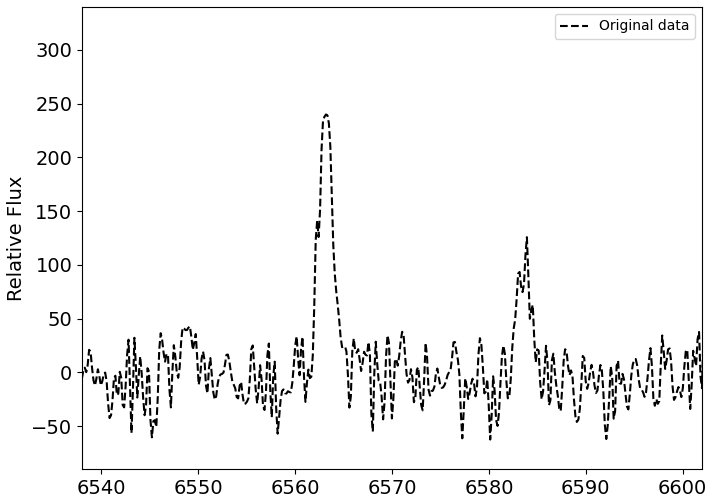}
\includegraphics[angle=0, width=7.5cm]{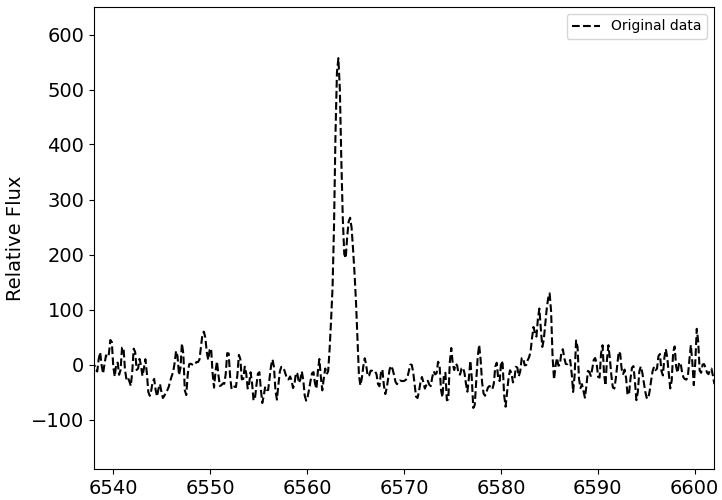}
\includegraphics[angle=0, width=7.5cm]{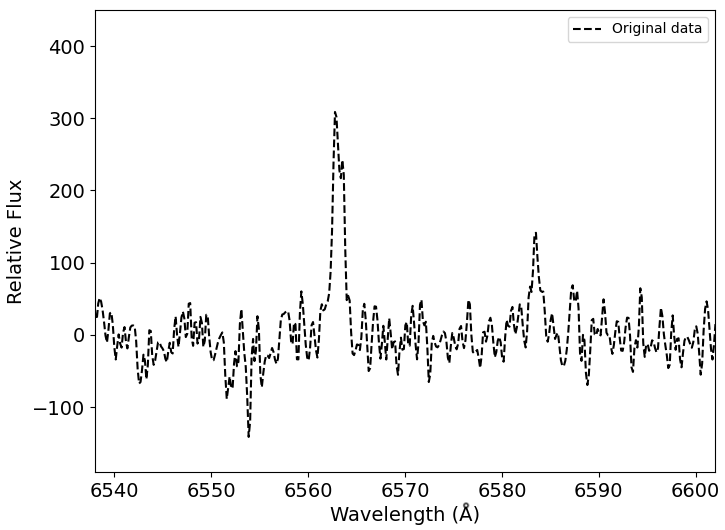}
\includegraphics[angle=0, width=7.5cm]{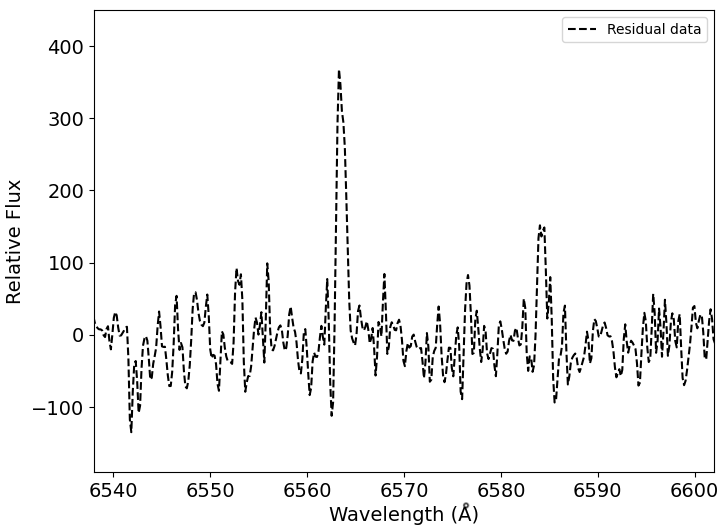}
\caption{LAMOST spectra ($6540-6600$ {\AA}) for P9, P10, P12, and P16 positions.}
\label{FigA2}
\end{figure*}

\begin{figure*}
\includegraphics[angle=0, width=7.5cm]{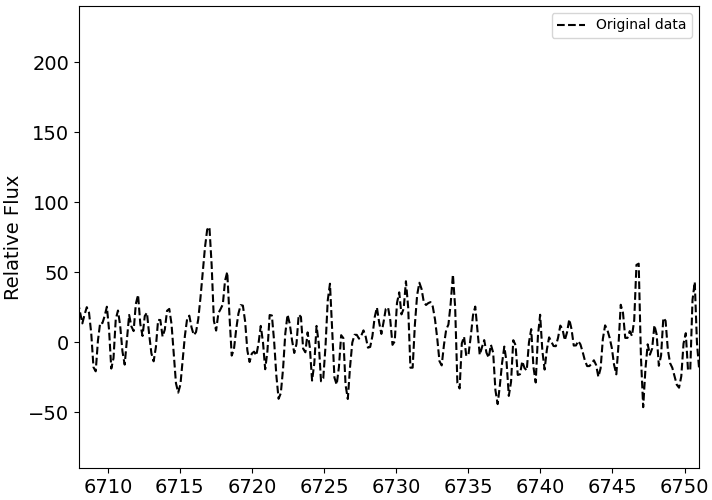}
\includegraphics[angle=0, width=7.5cm]{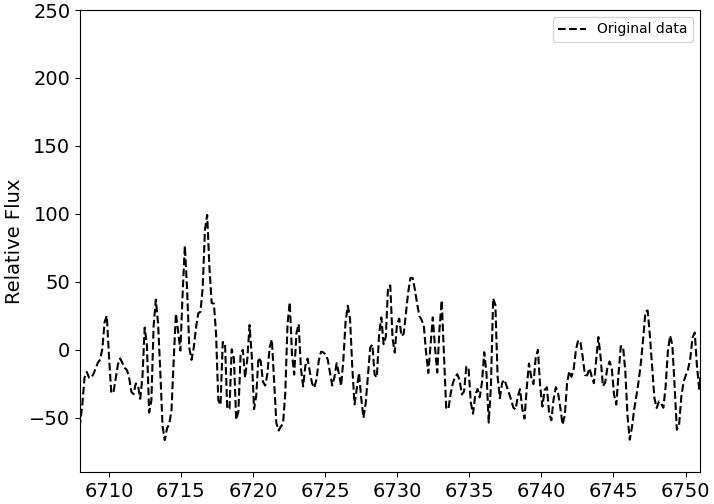}
\includegraphics[angle=0, width=7.5cm]{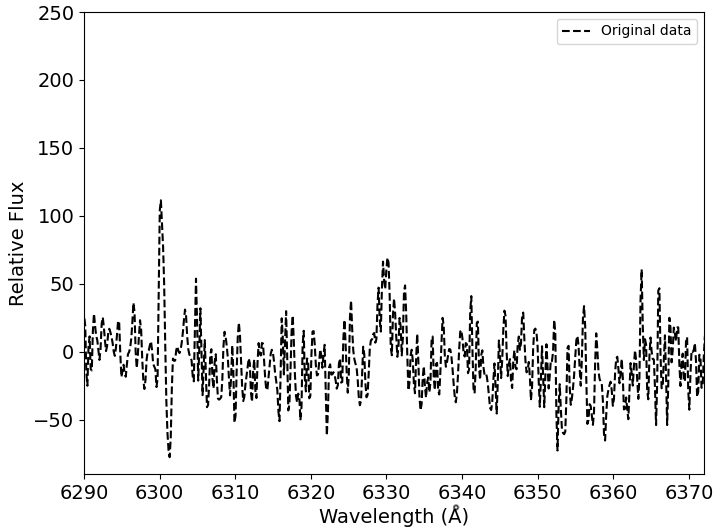}
\includegraphics[angle=0, width=7.5cm]{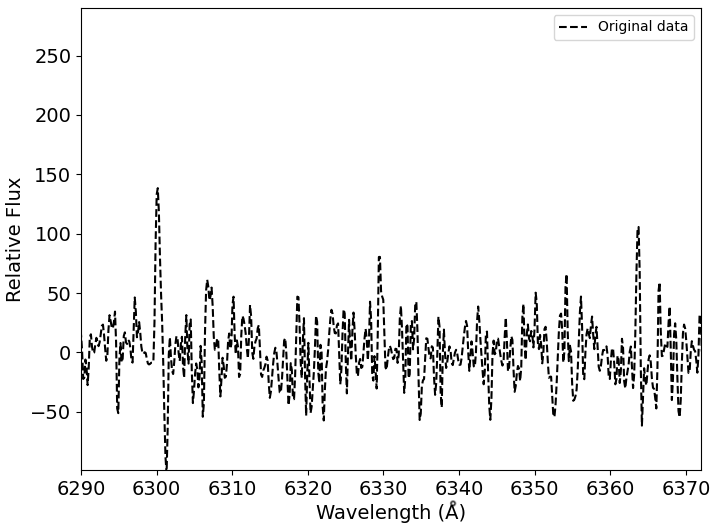}
\caption{LAMOST spectra ($6710-6750$ {\AA}) for P9 and P10 positions, and ($6290-6370$ {\AA}) for P10 and P16 positions.}
\label{FigA3}
\end{figure*}

In Figs~\ref{FigA1}-\ref{FigA3}, we show the LAMOST spectra.

\bsp	
\label{lastpage}
\end{document}